\documentclass[useAMS,usenatbib,fleqn]{mn2e}

\usepackage[draft]{hyperref}
\usepackage{amssymb}
\usepackage{amsmath}
\usepackage{mathtools}
\usepackage{enumitem}
\usepackage{graphicx}

\usepackage{nicefrac}
\usepackage{bigints}

\title[Polarization for SN~Ia explosion models]{Predicting polarization signatures for double-detonation and delayed-detonation models of Type Ia supernovae}
\author[M. Bulla et al.]{M.~Bulla,$^1$\thanks{E-mail: mbulla01@qub.ac.uk} S.~A.~Sim,$^{1,2}$ M.~Kromer,$^3$ I.~R. Seitenzahl,$^{2,4}$ M.~Fink,$^5$ \newauthor F.~Ciaraldi-Schoolmann,$^6$ F.~K.~R\"{o}pke,$^{7,8}$ W.~Hillebrandt,$^6$ R.~Pakmor,$^7$ \newauthor A.~J.~Ruiter$^{2,4}$ and S.~Taubenberger$^{6,9}$\\
$^1$Astrophysics Research Centre, School of Mathematics and Physics, Queen's University Belfast, Belfast BT7 1NN, UK\\
$^2$ARC Centre of Excellence for All-sky Astrophysics (CAASTRO)\\
$^3$The Oskar Klein Centre and Department of Astronomy, Stockholm University, AlbaNova, SE-106 91 Stockholm, Sweden\\
$^4$Research School of Astronomy and Astrophysics, Australian National University, Canberra, ACT 2611, Australia \\
$^5$Institut f\"{u}r Theoretische Physik und Astrophysik, Universit\"{a}t W\"{u}rzburg, Emil-Fischer-Stra{\ss}e 31, D-97074 W\"{u}rzburg, Germany\\
$^6$Max-Planck-Institut f\"{u}r Astrophysik, Karl-Schwarzschild-Str. 1, D-85748 Garching bei M\"{u}nchen, Germany \\
$^7$Heidelberger Institut f\"{u}r Theoretische Studien, Schloss-Wolfsbrunnenweg 35, D-69118 Heidelberg, Germany\\
$^8$Zentrum f{\"u}r Astronomie der Universit{\"a}t Heidelberg, Institut f{\"u}r Theoretische Astrophysik, Philosophenweg 12, D-69120 Heidelberg, Germany\\
$^9$European Southern Observatory, Karl-Schwarzschild-Str. 2, D-85748 Garching, Germany
}

\date{Accepted 2016 July 14. Received 2016 July 14; in original form 2016 April 21}

\usepackage{color}
\newcommand{\revised}[1]{\textcolor{black}{#1}}

\newcommand{\newCommandName}{pdf}  

\newcommand{\subch}{\mbox{D-DET }} 
\newcommand{\ch}{\mbox{N100-DDT }} 

\newcommand{\qrot}{\mbox{$Q_\text{ROT}$ }} 
\newcommand{\urot}{\mbox{$U_\text{ROT}$ }} 
\newcommand{\qrott}{\mbox{$Q_\text{ROT}$}} 
\newcommand{\urott}{\mbox{$U_\text{ROT}$}}

\begin{document}

\maketitle 

\begin{abstract}
Calculations of synthetic spectropolarimetry are one means to test multi-dimensional explosion models for Type Ia supernovae. In a recent paper, we demonstrated that the violent merger of a 1.1 and 0.9~M$_{\odot}$ white dwarf binary system is too asymmetric to explain the low polarization levels commonly observed in normal Type Ia supernovae. Here, we present polarization simulations for two alternative scenarios: the sub-Chandrasekhar mass double-detonation and the Chandrasekhar mass delayed-detonation model. Specifically, we study a two-dimensional double-detonation model and a three-dimensional delayed-detonation model, and calculate polarization spectra for multiple observer orientations in both cases. 
We find modest polarization levels ($<$~1~per~cent) for both explosion models. Polarization in the continuum peaks at $\sim$~0.1$-$0.3~per~cent and decreases after maximum light, in excellent agreement with spectropolarimetric data of normal Type Ia supernovae. Higher degrees of polarization are found across individual spectral lines. In particular, the synthetic Si\,{\sc ii}~$\lambda6355$ profiles are polarized at levels that match remarkably well the values observed in normal Type Ia supernovae, while the low degrees of polarization predicted across the O\,{\sc i}~$\lambda7774$ region are consistent with the non-detection of this feature in current data.
We conclude that our models can reproduce many of the characteristics of both flux and polarization spectra for well-studied Type Ia supernovae, such as SN~2001el and SN~2012fr. However, the two models considered here cannot account for the unusually high level of polarization observed in extreme cases such as SN~2004dt.
%
\end{abstract}

\section{Introduction}
\label{introduction}

Despite their relevance to cosmology \citep{riess1998,perlmutter1999}, Type Ia supernovae (SNe~Ia) are still poorly understood. While believed to stem from thermonuclear explosions of carbon-oxygen white dwarfs (WDs), answers to the questions of when, why and how these events are triggered remain unclear \citep{hillebrandt2013,maoz2014}. Recent explosion simulations have led to a better understanding of the physics involved, and comparisons of synthetic light curves and spectra with observations have played a key role in identifying which explosions scenarios are most promising. However, unambiguous discrimination between models is still challenging, even for the best observed nearby supernovae \citep{roepke2012}.

Polarization offers a unique opportunity to discriminate between the variety of possible explosion scenarios. The observational evidence that SNe~Ia are associated with rather low levels of polarization ($\lesssim$~1~per~cent) demands modest asphericities in the progenitor system and/or explosion mechanism, thus providing the means to effectively test different explosion models. Although predictions have been made using idealized geometries with simple departures from spherical symmetry -- such as ellipsoidal structures \citep{wang1997,howell2001,hoeflich2006,patat2012}, clumped  and toroidal shells \citep{kasen2003} or large scale asymmetries associated with ejecta overrunning the companion star \citep{kasen2004} -- until recently, no spectropolarimetric studies had been made for full multi-dimensional hydrodynamic explosion simulations. We have therefore started our \revised{long-term} project aiming to predict polarization signatures for \revised{a series of} modern SN~Ia hydrodynamic models. \revised{In the first paper of this series} \citep{bulla2016}, \revised{we have demonstrated the power of this approach for} the violent merger of two 1.1 and 0.9~M$_{\odot}$ WDs as presented by \citet{pakmor2012}. Despite matching luminosities and spectra of normal SNe~Ia reasonably well, this model was found to be too asymmetric to reproduce the polarization levels seen for the majority of normal SNe~Ia. 

\revised{In this second paper}, we carry out polarization spectral synthesis for examples of two alternative scenarios: the ``delayed-detonation" of a near Chandrasekhar mass (\mbox{M$_\text{ch}$}) WD and the ``double-detonation" of a sub-\mbox{M$_\text{ch}$} WD. 
In the ``delayed-detonation" model \citep[e.g.][]{khokhlov1991,hoeflich1995,plewa2004,roepke2007b,kasen2009,blondin2013,seitenzahl2013}, a WD is thought to accrete material from a non-degenerate companion star and explode following carbon ignition near the WD centre which occurs when the mass has grown close to \mbox{M$_\text{ch}$}. Initially, a carbon deflagration is ignited but delayed-detonation models posit that during the evolution of the explosion a detonation occurs. The second scenario we consider, the ``double-detonation" model
\citep[e.g.][]{nomoto1980,taam1980,livne1990a,shen2009,fink2010,moll2013}, describes the explosion of a sub-\mbox{M$_\text{ch}$} WD. In this model, the explosion is triggered by the detonation of a helium surface layer that has been accreted from a helium-rich companion star. The shock wave from this detonation then triggers a second detonation in the core.  Both delayed-detonation and double-detonation scenarios have received considerable attention over the past decade as they reproduce SN~Ia light curves and spectra reasonably well
\citep{hoeflich1995,hoeflich1996,kasen2009,kromer2010,roepke2012,sim2013,blondin2015} and could provide an important contribution to the SN~Ia population \citep[e.g.][]{hachisu2008,mennekens2010,ruiter2011}. 


In this paper we perform polarization calculations for one explosion model of \citet{fink2010} and one model of \citet{seitenzahl2013}. The former is chosen as it has been used in several studies (\citealt{scalzo2014a}a; \citealt{scalzo2014b}b; \citealt{childress2015}; \citealt{kosenko2015}) 
as a benchmark for the double-detonation scenario. The latter is selected since it has been widely used (\citealt{roepke2012}; \citealt{summa2013}; \citealt{scalzo2014a}a; \citealt{childress2015}; \citealt{fransson2015}; \citealt{kosenko2015}; \citealt{sneden2016}) as representative of a class of delayed-detonation models in which a deflagration-to-detonation transition (DDT) is assumed to occur spontaneously during the propagation of the deflagration. 

In Section~\ref{models} we summarize properties of the two specific explosion models, while in Section~\ref{simul} we discuss the details of our radiative transfer calculations. We present synthetic observables for both models in Section~\ref{synthobs} and comparisons with spectropolarimetric data of SNe~Ia in Section~\ref{compobs}. Finally, we discuss our results and draw conclusions in Section~\ref{conclusions}.



\section{Explosion Models}
\label{models}

\begin{figure*}
\begin{center}
\includegraphics[width=0.975\textwidth,clip=True,trim=-8pt 0pt 0pt 0pt]{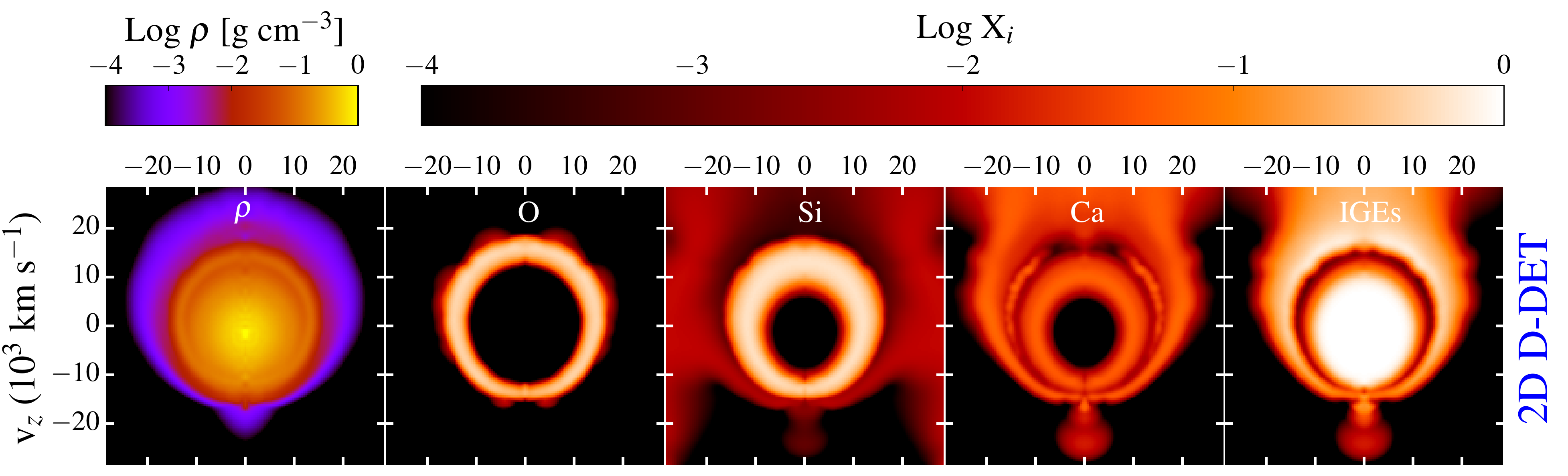}
\newline
\includegraphics[width=0.97\textwidth]{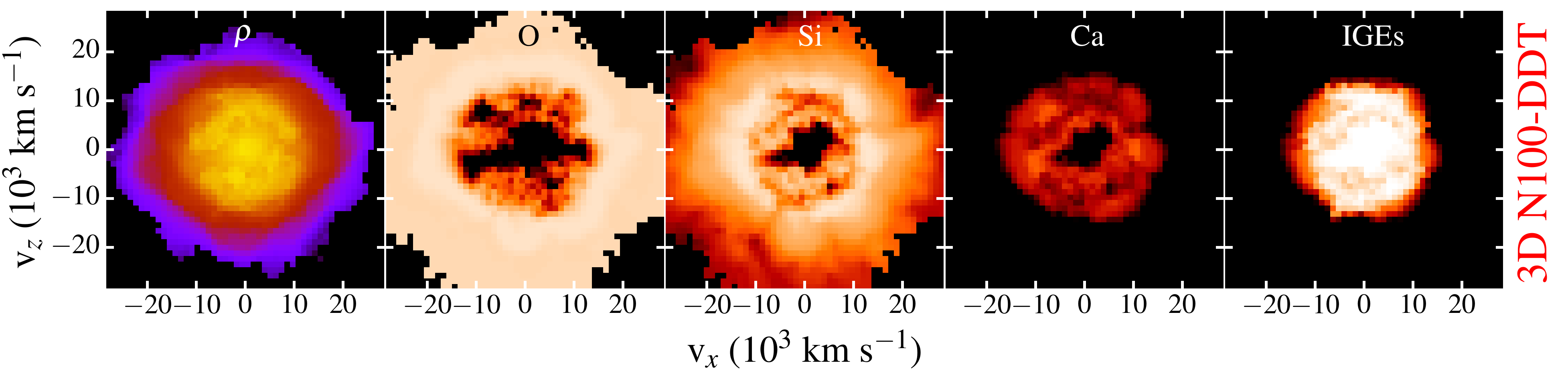}
\caption{Density and composition of the ejecta 100~s after explosion for the 2D \subch (upper panels) and 3D \ch (bottom panels) models studied in this work. The density $\rho$ and mass fractions X$_i$ of oxygen, silicon, calcium and IGEs (summed from scandium to zinc) are shown in the $x$-$z$ plane.}
\label{ejecta}
\end{center}
\end{figure*}

In this section we give a brief description of the two explosion models investigated in this study:
a double-detonation model (selected from the study by \citealt{fink2010}) and a delayed-detonation model (from \citealt{seitenzahl2013}).
\revised{We note that neither of the models we consider is perfect: in both cases, previous studies (see below) have identified shortcomings in matching observed spectra and light curve properties of normal SNe~Ia. Nevertheless, we adopt these two example models as the most promising from the Fink et al. and Seitenzahl et al. series, respectively, and will focus on the added value of synthetic spectropolarimetry in understanding the strengths and weaknesses of their respective explosion scenarios.}

\subsection{Sub-M$_\text{ch}$ double-detonation model}
\label{subchmodel}

The six double-detonation models investigated by \citet{fink2010} comprise WDs with carbon-oxygen 
core masses increasing from 0.81 to 1.38~M$_{\odot}$ and helium-shell masses decreasing from 0.13 to 3.5$\times$10$^{-3}$~M$_{\odot}$. The specific helium-shell masses were selected to meet minimum conditions required for a helium detonation suggested by \citet{bildsten2007}. Hydrodynamic simulations were performed in two-dimensions (2D) adopting azimuthal symmetry about the $z$ axis. In all the models, the first detonation is ignited at a single point on the positive $z$ axis at the interface between the core and the helium shell. While the detonation wave sweeps along the core-shell interface, a shock wave moves through the core and converges at an off-centre point on the negative $z$ axis \citep{livne1990b}. A second detonation is then initiated around this point if temperature and pressure conditions are sufficiently high \citep{niemeyer1997,roepke2007a}.
\revised{We note that, although this model is only 2D, neither the adopted WD progenitor model nor the ignition condition break axi-symmetry and thus it is expected to be a good approximation that azimuthal symmetry is preserved.}

For our spectropolarimetric study, we select `model~3' of \citet{fink2010} since this configuration \revised{produces an amount of $^{56}$Ni (0.55~M$_{\odot}$) in the range observed for normal SNe~Ia (\citealt{stritzinger2006}; \citealt{scalzo2014a}a; \citealt{childress2015})} and is found to predict observables in \revised{reasonable} -- although not perfect -- agreement with normal SNe~Ia \citep{kromer2010}. This model (referred to as the \subch model in the rest of the paper) has a carbon-oxygen core of 1.025~M$_{\odot}$ and a helium shell of 0.055~M$_{\odot}$. Owing to the low densities in the \textit{shell}, 60~per~cent of the helium is left unburned and much of the material that is burned does not reach nuclear statistical-equilibrium (NSE). Specifically, the conditions in the model favour burning products rich in iron-group-elements (IGEs) lighter than $^{56}$Ni, such as $^{52}$Fe (11~per~cent), $^{48}$Cr (8~per~cent) and $^{44}$Ti (6~per~cent). Aside from modest fractions of $^{40}$Ca (4~per~cent) and $^{36}$Ar (4~per~cent), very few intermediate-mass elements (IMEs) are produced in the shell. 
In contrast, the \textit{core} detonation synthesizes 0.55~M$_{\odot}$ of $^{56}$Ni (96~per~cent of the total IGE production in the core) and 0.37~M$_{\odot}$ of IMEs. 

The upper panels in Fig.~\ref{ejecta} show the density and composition of the ejecta when the system has already entered the homologous expansion phase ($\sim$~8~s after explosion). Due to the asymmetries introduced by the explosion mechanism, the ejecta structure is fairly aspherical\footnote{We note that the specific ejecta morphology depends on the ignition geometry. Asymmetries produced in the \subch model of \citet{fink2010} follow from the choice of a one-point ignition on the $z$-axis, while more symmetric ejecta are expected for 
more symmetric shell ignitions (see \citealt{fink2007}).
}. 
First, since the helium detonation is ignited in a point on the positive $z$-axis, ejecta from the \textit{shell} (e.g. the outer layer of calcium) are distributed over a wider range of velocities in the northern compared to the southern hemisphere. Second, the off-center ignition of the \textit{core} detonation causes the northern regions of the core to be less compressed than those on the opposite side. This has the effect of producing IMEs (e.g. silicon) that are more abundant and extend over a wider velocity range in the northern compared to the southern hemisphere (see also section~4.3 of \citealt{fink2010}). 

\subsection{M$_\text{ch}$ DDT model}
\label{chmodel}

The set of three-dimensional (3D) DDT models studied by \citet{seitenzahl2013} comprises WDs with mass close to M$_\text{ch}$.
Twelve of the fourteen models differ only in ignition geometry of the deflagration (parametrized in terms of a number of spherical ignition kernels, ranging from 1 to 1600). The remaining two models are alternative versions of the 100-kernel model that consider different WD central densities. For all the models, the kernels were placed near the centre of the WD. Following the ignition, the deflagration flame propagates and turbulence develops. Once the turbulent velocity fluctuations at the flame front become sufficiently strong \citep{ciaraldi2013,seitenzahl2013}, a transition to a supersonic detonation is invoked, which totally disrupts the WD. 


For our spectropolarimetric study, we select the `N100' model of \citet{seitenzahl2013}. This model (hereafter referred to as \ch model) is chosen as it yields an amount of $^{56}$Ni (0.6M$_{\odot}$) consistent with the values derived for the bulk of normal SNe~Ia (\citealt{stritzinger2006}; \citealt{scalzo2014a}a; \citealt{childress2015}) and because it provides a good match to observed spectra and light curves \citep{roepke2012,sim2013}. This specific configuration produces 0.84~M$_{\odot}$ of IGEs and 0.45~M$_{\odot}$ of IMEs.

The lower panels of Fig.~\ref{ejecta} show the density and composition of the ejecta for this model 100~s after explosion. The rather symmetric ignition geometry of this specific explosion simulation results in an overall spherical distribution of the yields. However, there are clear small-scale asymmetries in the oxygen and IME distributions that can be expected to imprint observable signatures. 

\section{Simulations}
\label{simul}

\subsection{Radiative transfer calculations}
\label{radtransf}
To extract synthetic observables for the explosion models presented in Section~\ref{models}, we carried out calculations using our 3D Monte Carlo (MC) radiative transfer code \textsc{artis} \citep{sim2007,kromer2009,bulla2015}. Radiative transfer calculations for the same \subch and \ch models were already performed by \citet{kromer2010} and \citet{sim2013}, respectively, but here we present new simulations that include polarization. We first remap each explosion model to a Cartesian grid: in line with the previous studies, we remap the 2D \subch model to a $100^3$ grid and the 3D \ch model to a $50^3$ grid. We then follow the propagation of $N_\text{q}$ MC quanta from 2 to 120~d after explosion using a series of 111 logarithmically spaced time-steps. We assume local thermodynamic equilibrium for the first 10 time-steps ($t<2.95$\,d) and a grey approximation for optically thick cells \citep{kromer2009}. Compared to \citet{kromer2010} and \citet{sim2013}, here we use a more extended atomic data set \citep{gall2012} that includes about \revised{$8.6\times10^6$} lines and ions \textsc{i}~-~\textsc{vii} for \mbox{20~\textless~Z \textless~29}. Moreover, in order to achieve low MC noise levels in the polarization spectra (see also Section~\ref{noise}), we utilize a larger number of MC quanta, $N_\text{q}=5.12\times10^8$, for both the \subch \citep[cf. with $N_\text{q}=2\times10^7$ of][]{kromer2010} and the \ch\citep[cf. with $N_\text{q}=1.024\times10^8$ of][]{sim2013} model.

\begin{figure*}
\begin{center}
\includegraphics[width=0.282\textwidth]{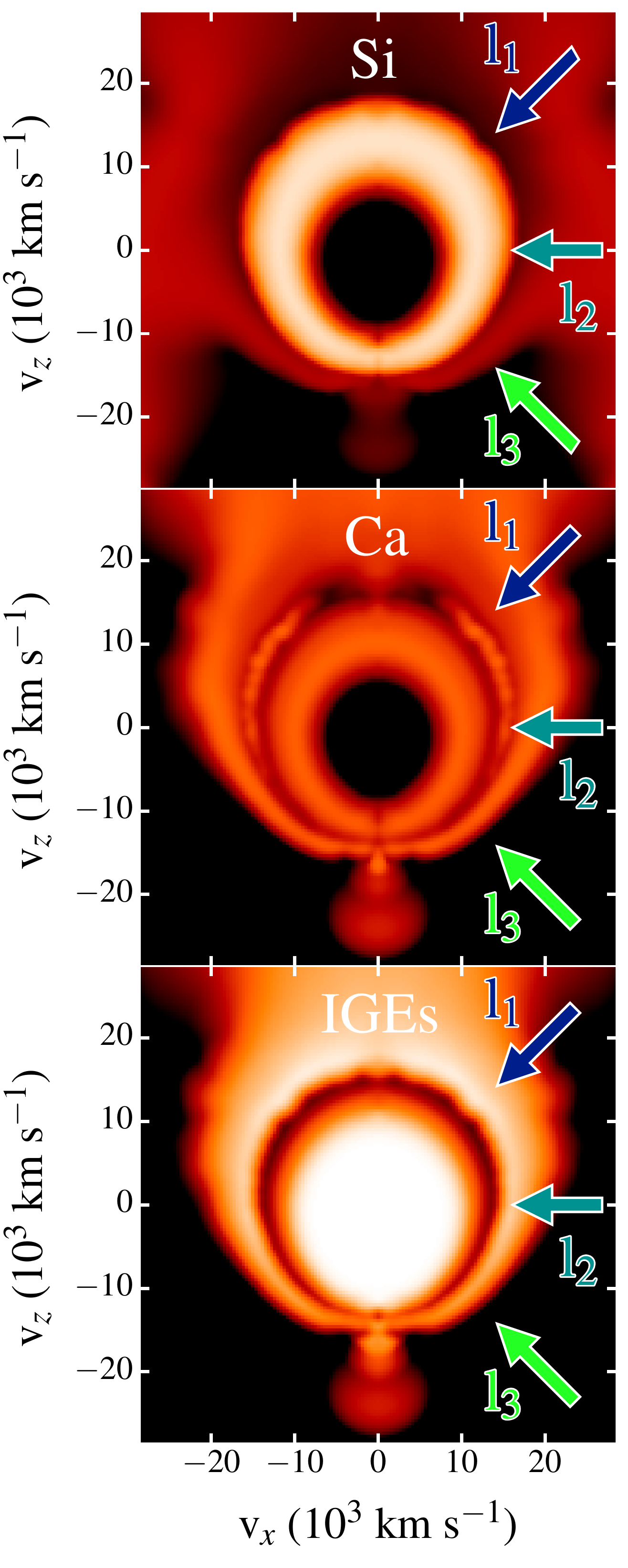}
\includegraphics[width=0.71\textwidth]{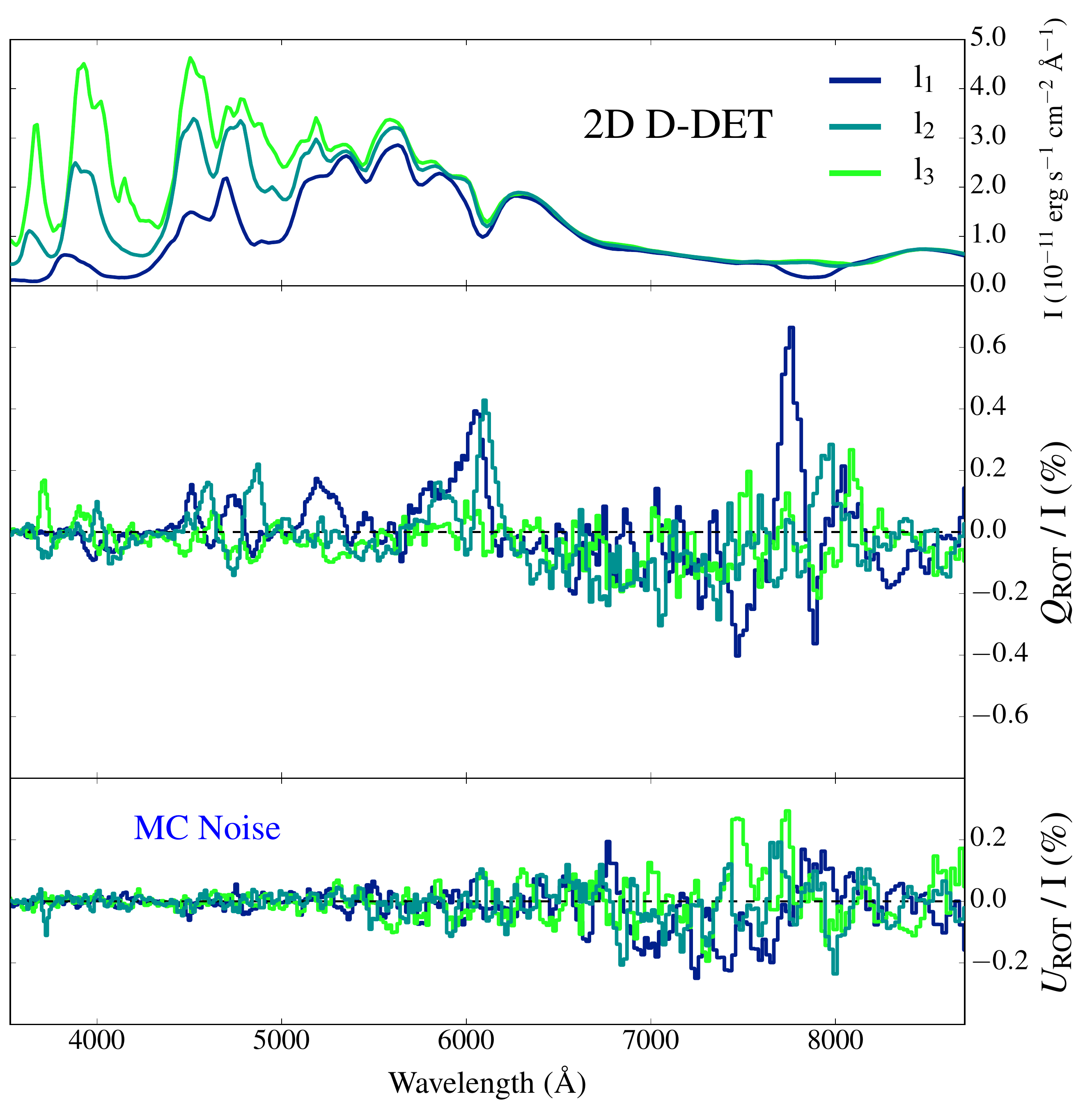}
\caption{\textit{Left-hand panels.} Orientations of the three observers selected for the \subch model ($\bmath{l_1}$, $\bmath{l_2}$ and $\bmath{l_3}$) with respect to the ejecta composition. The mass fractions of silicon (top), calcium (middle) and IGEs (bottom) are represented using the same colour scale of Fig.~\ref{ejecta}. \textit{Right-hand panels.} Flux (top) and polarization \qrot (middle) spectra at 18.5~d after explosion ($B-$band maximum light) for the three chosen viewing angles. Polarization \urot spectra are reported in the bottom panel and provide proxies for the MC noise levels in the \qrot spectra (see discussion in Section~\ref{noise}). Polarization spectra are Savitzky-Golay filtered \revised{using a first-order polynomial and a window} of 3 pixels ($\sim$~50~\AA) for clarity. The model flux is given for a distance of $1$~Mpc.}
\label{specmax_subch}
\end{center}
\end{figure*}

\begin{figure*}
\begin{center}
\includegraphics[width=0.71\textwidth]{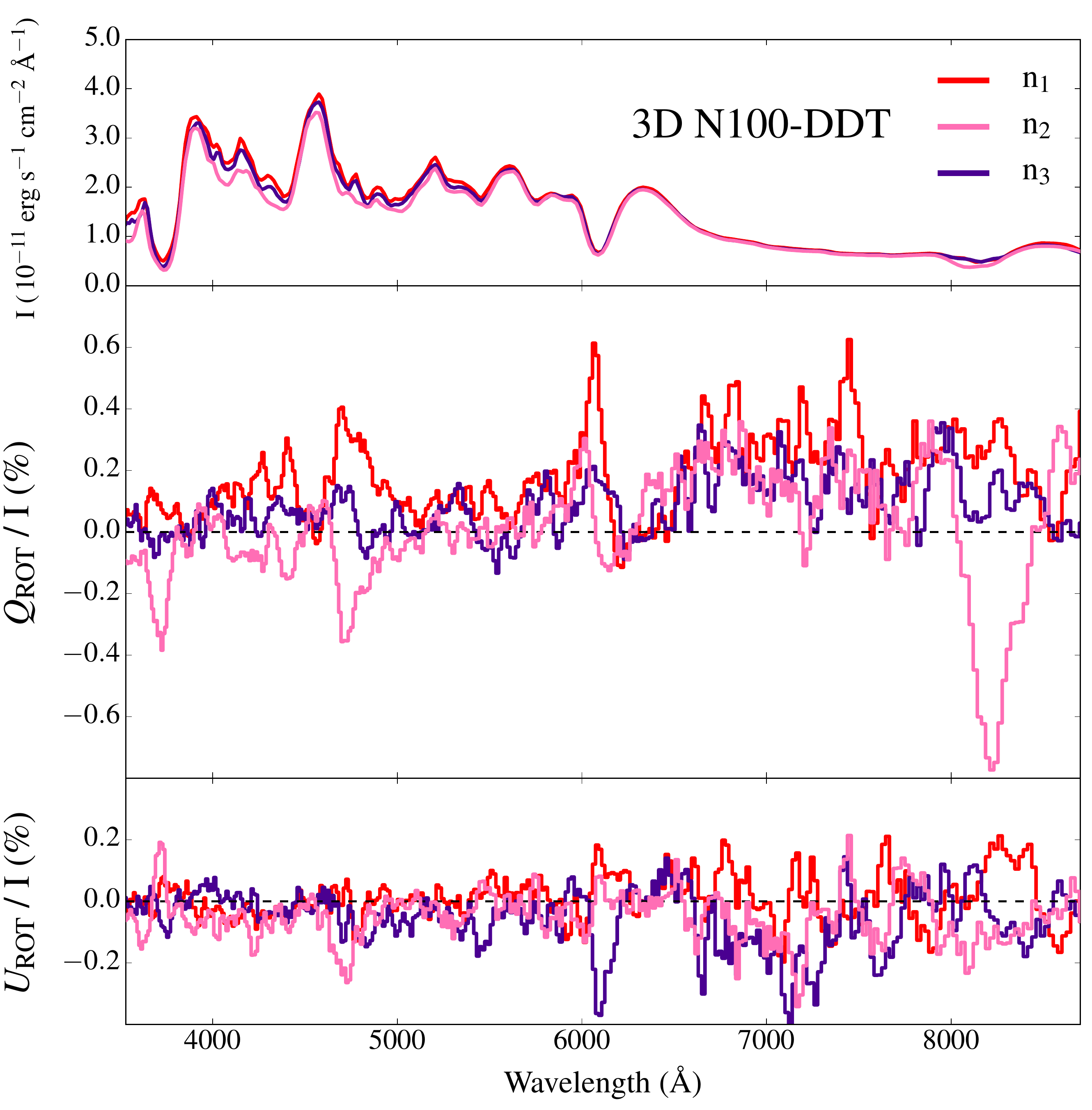}
\includegraphics[width=0.282\textwidth]{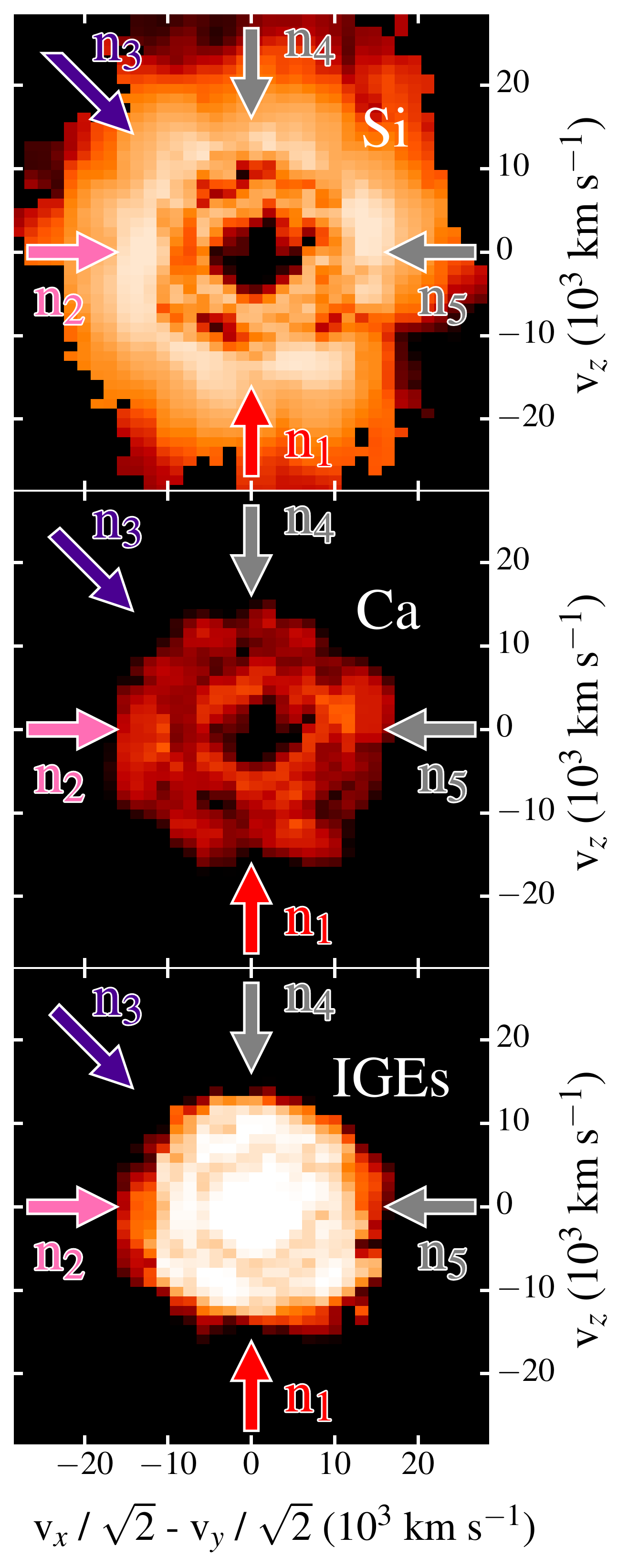}
\caption{\textit{Left-hand panels.} Flux (top) and polarization \qrot (middle) spectra extracted for the \ch model at 17~d after explosion ($B-$band maximum light) along the $\bmath{n_1}$, $\bmath{n_2}$ and $\bmath{n_3}$ orientations. Polarization \urot spectra are reported in the bottom panel and reflect deviations from a single-axis geometry (see discussion in Section~\ref{noise}). \urot spectra in the \subch model (see bottom panel of Fig.~\ref{specmax_subch}) provide estimates of the MC noise levels in both \qrot and \urott. Polarization spectra are Savitzky-Golay filtered \revised{using a first-order polynomial and a window} of 3 pixels ($\sim$~50~\AA) for clarity. The model flux is given for a distance of $1$~Mpc. \textit{Right-hand panels.} Orientations of the five observers selected for the \ch model ($\bmath{n_1}-\bmath{n_5}$) with respect to the ejecta composition. The mass fractions of silicon (top), calcium (middle) and IGEs (bottom) are represented using the same colour scale of Fig.~\ref{ejecta}.}
\label{specmax_ddt}
\end{center}
\end{figure*}

Following the technique described by \citet{bulla2015}, we extract high signal-to-noise synthetic observables for specific observer orientations. For the 2D \subch model, we select three orientations in the $x$-$z$ plane\footnote{The choice of restricting our study to the $x$-$z$ plane is sufficient since the model has azimuthal symmetry about the $z$ axis.} so that the ejecta asymmetries are properly sampled from north to south pole (see left panels of Fig.~\ref{specmax_subch}): $\bmath{l_1}=(\nicefrac{1}{\sqrt{2}},0,\nicefrac{1}{\sqrt{2}})$, $\bmath{l_2}=(1,0,0)$, $\bmath{l_3}=(\nicefrac{1}{\sqrt{2}},0,-\nicefrac{1}{\sqrt{2}})$. For the 3D \ch model, we select five observer orientations (see right panels of Fig.~\ref{specmax_ddt}). Guided by the findings of \citet{sim2013}, we first choose the orientations for which the model is brightest, $\bmath{n_1}=(0,0,-1)$, and faintest, $\bmath{n_2}=(-\nicefrac{1}{\sqrt{2}},\nicefrac{1}{\sqrt{2}},0)$, and then select three additional directions to sample the plane defined by $\bmath{n_1}$ and $\bmath{n_2}$: $\bmath{n_3}=(-\nicefrac{1}{2},\nicefrac{1}{2},\nicefrac{1}{\sqrt{2}})$, $\bmath{n_4}=(0,0,1)$ and $\bmath{n_5}=(\nicefrac{1}{\sqrt{2}},-\nicefrac{1}{\sqrt{2}},0)$. Flux and polarization spectra are extracted for each orientation between 10 and 30~d after explosion in the wavelength range 3500$-$10\,000~\AA. To decrease the MC noise levels in the red regions of the polarization spectra -- where the flux is lower -- we perform additional simulations with $N_\text{q}=2\times10^8$ and restricting to wavelengths larger than 5800~\AA{} \citep[see][]{bulla2015}.

To map out the range of polarization covered by the explosion models, we also carried out two calculations with fewer packets ($N_\text{q}=1.5\times10^8$) and with polarization spectra extracted for an additional number of viewing angles. Specifically, 17 and 21 extra-orientations were selected to properly sample the ejecta of the 2D \subch and 3D \ch model, respectively. Results of these simulations will be presented in Section~\ref{polsilicon}.

\subsection{Polarization decomposition and MC noise}
\label{noise}

An important drawback of MC simulations is their stochastic nature: MC noise in our extracted spectra can become sufficiently large to make the identification of real features challenging. This is particularly critical for the explosion models presented in this study since the modest asphericities of their ejecta (see Fig.~\ref{ejecta} and discussion in Section~\ref{models}) are expected to yield low values of polarization ($\lesssim$~1~per~cent). Here we outline the choices we make to present our results and discriminate between intrinsic polarization signals and MC noise in this paper.

Polarization is sometimes represented in terms of the Stokes parameters $Q$ and $U$, sometimes in terms of degree of polarization $P=\sqrt{Q^2+U^2}/I$ and polarization angle $\chi$ (derived from $\tan2\chi$~=~$U/Q$). When dealing with low signal-to-noise levels, however, the former representation is usually preferred as $P$ is known to suffer from the so-called polarization bias \citep[see e.g.][and references therein]{patat2006}. Specifically, every contribution to $Q$ and $U$ in our simulations will be distributed around the true value $\bar{Q}$ and $\bar{U}$ because of MC noise fluctuations. In the presence of small signals, $Q$ and $U$ contributions can assume both positive and negative values. Given that the polarization percentage $P$ is defined positive, however, negative $Q$ and $U$ contributions will add only in the positive direction leading to a systematic overestimate of the true value $\bar{P}$. 

Therefore, in this paper we will initially work in terms of the Stokes parameters as these do not suffer from the polarization bias. Following \citet{leonard2001} and \citet{wang2003}, we also introduce a new set of Stokes parameters \qrot and \urot by rotating $Q$ and $U$ by a given angle $\alpha$ that brings the strongest polarization signal along \qrott:
\begin{equation}
\label{eq_rotation}
\begin{cases} \qrott = Q \cos\alpha -  U\sin\alpha \\
\urott = Q \sin\alpha +  U\cos\alpha ~~~.\\
\end{cases}
\end{equation}
Specifically:
\begin{itemize}[leftmargin=*]
\item for the 2D \subch model, we exploit the axial symmetry of the explosion and plot $Q$ as \qrot and $U$ as \urot (i.e. $\alpha=0$). Thanks to the azimuthal symmetry around the $z$-axis, \qrot is expected to contain \textit{all} the polarization signal, while \urot should be consistent with zero (within the MC noise level, see below);
\item for the 3D \ch model, we identify $\alpha$ via a weighted least-squared fitting of a straight line in the $Q/U$ plane. After the rotation, \qrot is expected to contain \textit{most of} the polarization signal, while \urot to be reflective of deviations from a single-axis geometry \citep[see also][]{wang2008}.
\end{itemize}
We will, however, prefer to use $P$ when discussing polarization light curves in Section~\ref{polevol} (as the polarization bias is much less important there) and when comparing our models with data in Section~\ref{compobs} (as the comparison is more straightforward in terms of absolute percentage levels and this representation has been widely used in previous studies).  

In this paper we estimate MC noise levels by exploiting the axial symmetry of the 2D \subch model. Owing to the symmetry about the $z$-axis, no intrinsic polarization is expected in the \urot spectrum for this model. Deviations from zero in the \urot spectrum are instead associated with statistical fluctuations and can be used as a convenient proxy for the MC noise in the \qrot spectrum. Given that both explosion models are investigated with the same number of MC quanta $N_\text{q}$, we assume their statistical fluctuations to be comparable. 
Therefore, we also consider the \urot spectra for the 2D \subch model as representative of the MC noise in the \qrot and \urot spectra of the 3D \ch model. 

\section{Synthetic observables}
\label{synthobs}

\begin{figure}
\begin{center}
\includegraphics[width=0.48\textwidth,clip=True,trim=5pt 0pt 0pt 0pt]{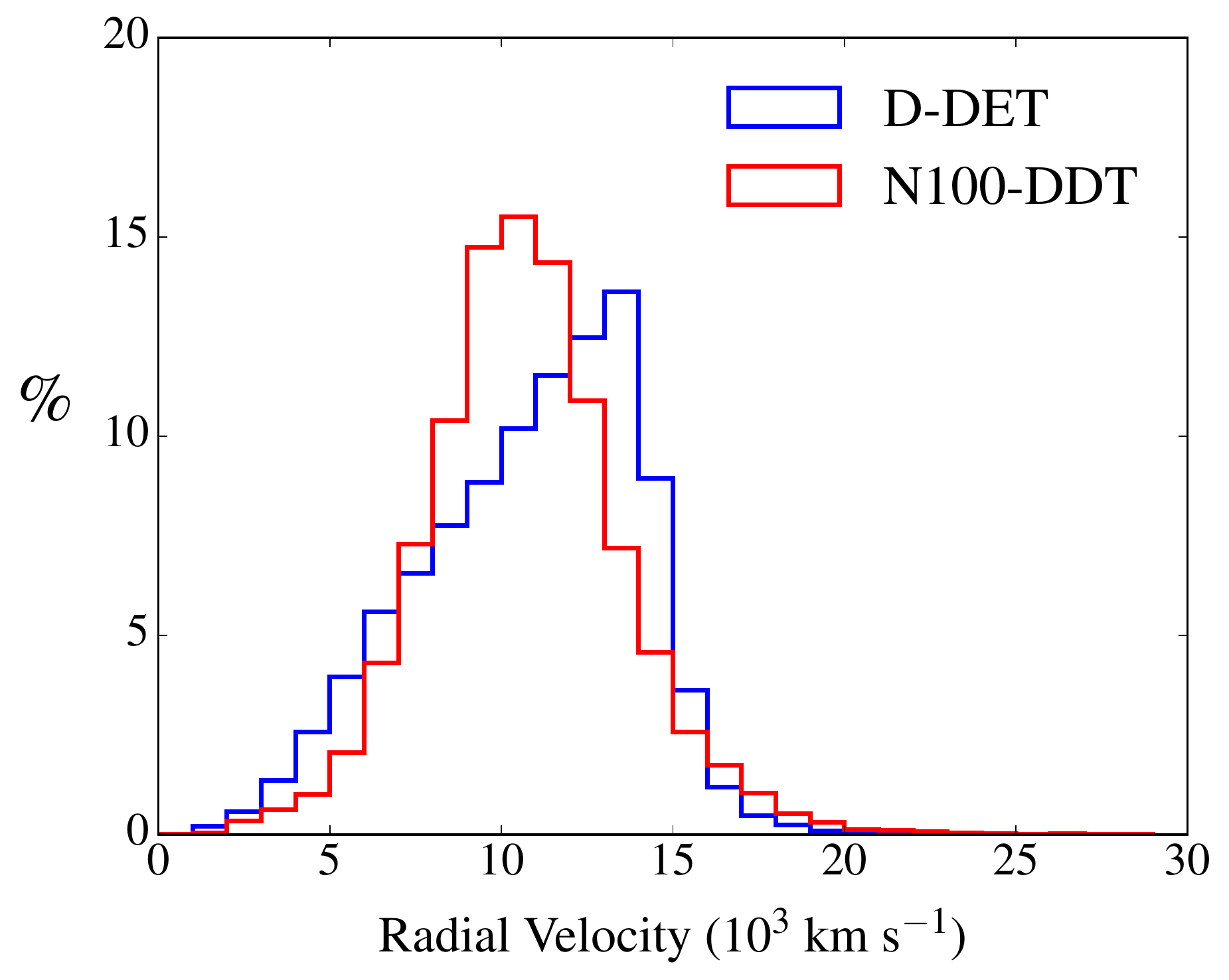}
\caption{Region of last scattering for polarizing contributions. Distributions include radial velocities at the points where packets underwent a polarizing interaction (i.e. electron scattering) as last interaction before escaping towards the observer. 10$^6$ packets have been used for both the \subch (blue) and the \ch (red) model.}
\label{lastscatt}
\end{center}
\end{figure}

In this section, we present synthetic observables for the \subch and \ch models introduced in Section \ref{models}. For the \ch model, qualitatively similar conclusions are found for the five different orientations and thus we present results for only three viewing angles ($\bmath{n_1}-\bmath{n_3}$) for simplicity; the calculations for the other two observer orientations will be included in the comparison with data in Section~\ref{compobs}, however. We discuss flux and polarization spectra around maximum light in Section~\ref{polmax}\revised{, illustrate how asymmetries in the element distribution are linked to specific polarization features in Section~\ref{geometry}} and finally study the temporal evolution of the polarization signal in Section~\ref{polevol}.

\subsection{Polarization around maximum light}
\label{polmax}

\subsubsection{2D \subch model} 
\label{subchmax}

In Fig.~\ref{specmax_subch} we show flux and polarization spectra for the selected \subch model orientations $\bmath{l_1}$ (north), $\bmath{l_2}$ (equator) and $\bmath{l_3}$ (south). Spectra are reported at t$^\text{B}_\text{max}=$~18.5~d after explosion, that is around $B-$band maximum light for all the observers. As described by \citet{kromer2010}, the asymmetries in the ejecta (see Fig.~\ref{ejecta}) translate into strongly viewing-angle dependent flux spectra. Compared to packets escaping in the southern hemisphere (along $\bmath{l_3}$), those on the opposite side (along $\bmath{l_1}$) have to travel through a more extended layer of IGEs and are therefore more likely to be absorbed and reemitted at longer wavelengths. This effect leads to a much redder spectrum for the orientation in the northern hemisphere. Owing to the off-center ignition of the core (see Section \ref{models}), IMEs are also more abundant and distributed over a wider velocity range in the northern hemisphere. IME spectral features extracted along $\bmath{l_1}$ are thus stronger and broader compared to those along $\bmath{l_3}$ (see for instance the Si\,{\sc ii}~$\lambda6355$ line and Ca\,{\sc ii} IR triplet). Intermediate spectral properties are found for the equatorial observer orientation ($\bmath{l_2}$). 

Polarization levels extracted for the three chosen observer orientations are relatively low ($\left|\qrott\right|$~$\lesssim$~0.7~per~cent). Polarization signals throughout the spectrum are generally stronger for the fainter orientation ($\bmath{l_1}$) and weaker for the brighter orientation ($\bmath{l_3}$), although deviations from this behaviour are seen. We do, however, find similar degrees of polarization (\qrott~$\sim$~$-$0.1~per~cent, see also Section~\ref{polevol}) for all observer orientations in the ``pseudo-continuum'' region 6500$-$7500~\AA, a region usually assumed to be devoid of strong lines and thus reflective of the continuum level \citep{kasen2004,leonard2005,patat2009}. Non-zero continuum polarization in SNe is representative of asymmetries in the underlying electron-scattering photosphere. Therefore, the extremely low levels found across the pseudo-continuum region 6500$-$7500~\AA{} arise as a direct consequence of the overall spherical symmetry of the inner regions of the ejecta (velocities between $\sim$~5000 and 15\,000~km~s$^{-1}$, see Fig.~\ref{ejecta}), from which most of the electron-scattered polarizing contributions originate (see~ Fig.~\ref{lastscatt}). 

Despite the small signal in the pseudo-continuum, the polarization increases at wavelengths corresponding to the absorption troughs of spectral lines. In the upper panels of Fig.~\ref{zoom}, we report polarization spectra between 5000 and 8500~\AA{} extracted along the $\bmath{l_1}$ direction. Matches between spectral lines and polarization peaks are unambiguous above 5000~\AA{}. The strongest polarization feature in this range is a peak of about 0.7~per~cent around 7700~\AA, which clearly corresponds to the Ca\,{\sc ii} IR triplet at high-velocities (v~$\sim$~$-$26\,000~km~s$^{-1}$) and reflects the asymmetric distribution of calcium in the outer shell \revised{(but see also discussion in Section~\ref{geometry_subch})}. Other peaks are instead found at lower velocities (between $-$10\,000 and $-$13\,000~km~s$^{-1}$) and attributed to silicon and sulphur transitions in the outer layers of the core detonation. Compared to calcium, these elements are more spherically distributed and thus characterized by weaker polarization signatures ($\left|\qrott\right|$~$\lesssim$~0.4~per~cent). Specifically, we find clear polarization signals for the Si\,{\sc ii}~$\lambda6355$ line (see also Section~\ref{polsilicon}), while smaller -- and possibly null -- signals are predicted for the \revised{S\,{\sc ii}~$\lambda5454$, S\,{\sc ii}~$\lambda5640$ and} Si\,{\sc ii}~$\lambda5972$ \revised{lines}. \revised{A clear polarization feature is also seen around 7500~\AA{}, with negative \qrot (i.e. 90$^\circ$ rotated from the rest of the features). In our simulations, this features can be attributed to a blend of O\,{\sc i}~$\lambda\lambda7772,7774,7775$ (hereafter O\,{\sc i}~$\lambda7774$), Si\,{\sc ii}~$\lambda\lambda7849,7850$ (Si\,{\sc ii}~$\lambda7849$) and Mg\,{\sc ii}~$\lambda\lambda7877,7896$ (Mg\,{\sc ii}~$\lambda7887$), with relative contributions that depend on orientation and epoch. The nature of this polarization spike and its 90$^\circ$ rotation from the rest of the features will be investigated in Section~\ref{geometry}.} With the exception of a polarization peak around 3700~\AA, associated with the Si\,{\sc ii}~$\lambda3859$ and/or Ca~H~and~K features, identifications to individual transitions are not obvious below $\sim$~5000~\AA{} due to the strong line blending suffered by these blue regions.

\subsubsection{3D \ch model}
\label{chmax}

In the top panel of Fig.~\ref{specmax_ddt} we show flux spectra for the \ch model orientations $\bmath{n_1}$, $\bmath{n_2}$ and $\bmath{n_3}$ at t$^\text{B}_\text{max}=$~17~d after explosion (around $B-$band maximum light for the three orientations). As shown in Fig.~\ref{ejecta}, the ejecta distribution in the \ch model is rather spherical and does not have IGE material at very high velocities, in contrast to the \subch model. Compared to the \subch model, the \ch model is therefore characterized by a much smaller viewing angle dependence in the flux spectra \citep{sim2013}. One noticeable difference is found for the blue-shifted absorption component of the Ca\,{\sc ii} IR triplet, which is stronger when observed from an equatorial orientation $\bmath{n_2}$. This can be understood by noting that the calcium distribution is not perfectly spherical (see right panels of Fig.~\ref{specmax_ddt}) and offers higher opacities to packets escaping towards $\bmath{n_2}$ compared to those escaping in the other two directions considered here.

\begin{figure*}
\begin{center}
\hspace{0.65cm}
\includegraphics[width=0.424\textwidth,clip=True,trim=0pt 0pt 0pt 0pt]{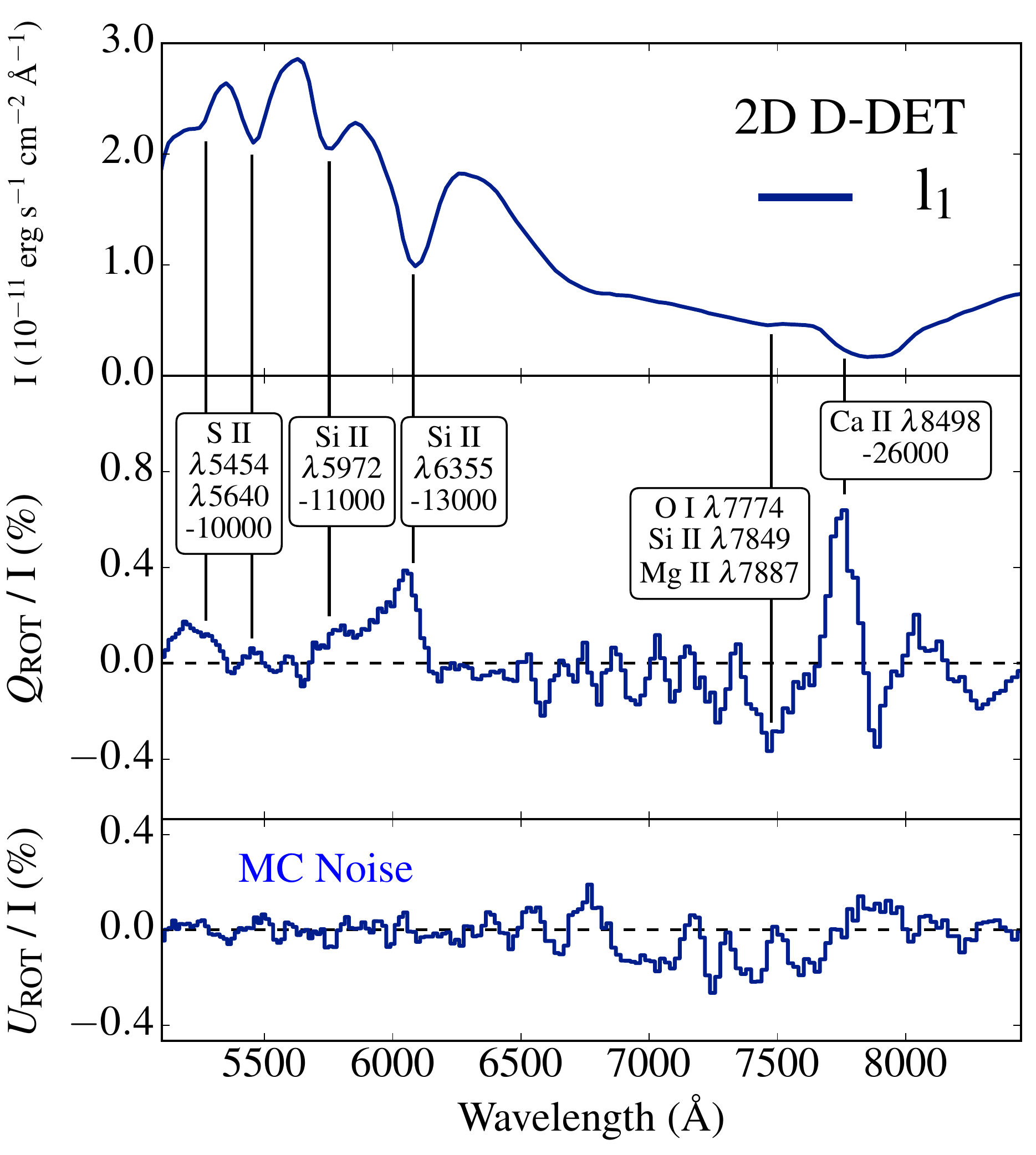}\includegraphics[width=0.4\textwidth,clip=True,trim=0pt 0pt 0pt 0pt]{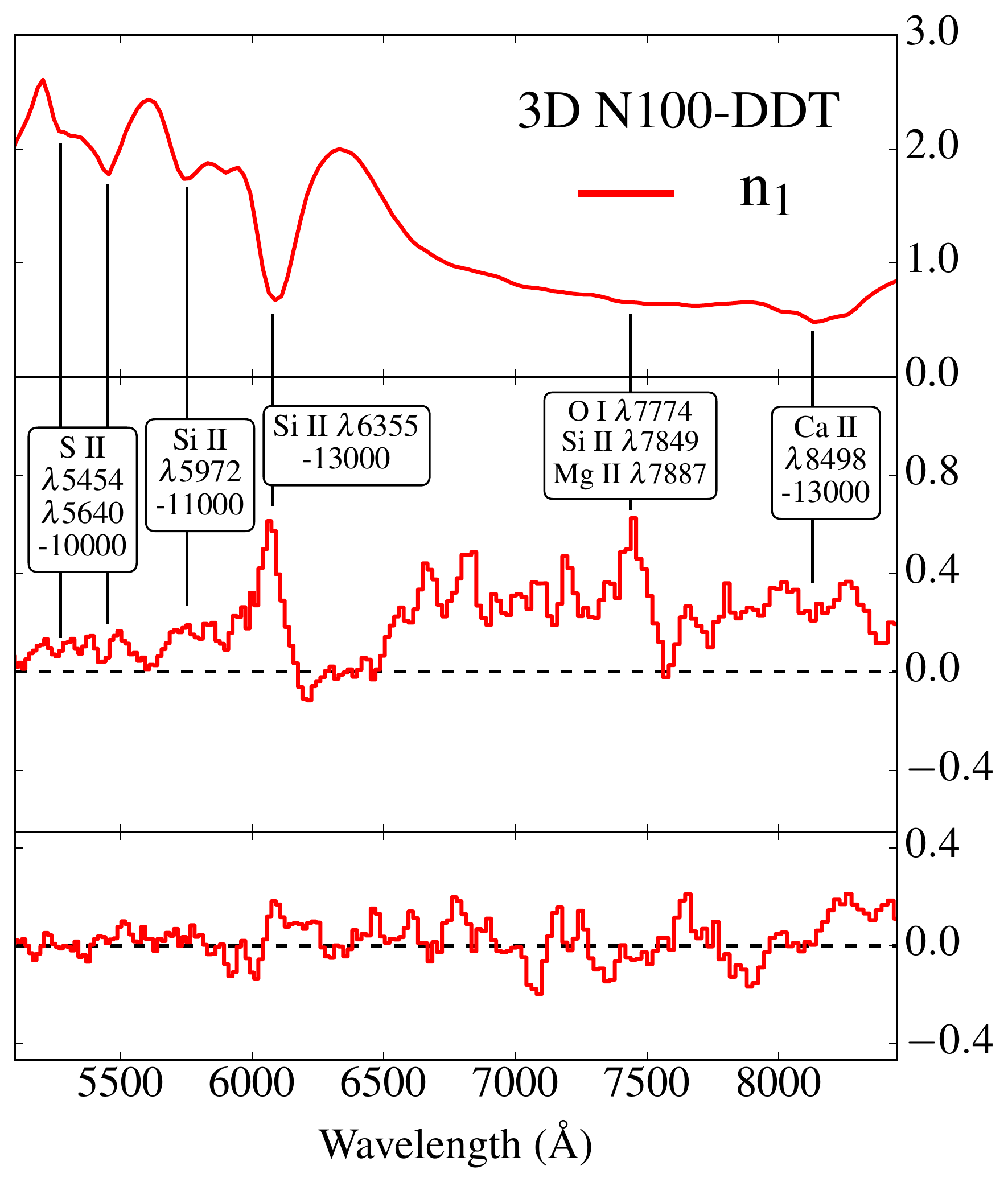}
\newline
\includegraphics[width=0.417\textwidth,clip=True,trim=0pt 0pt 0pt -15pt]{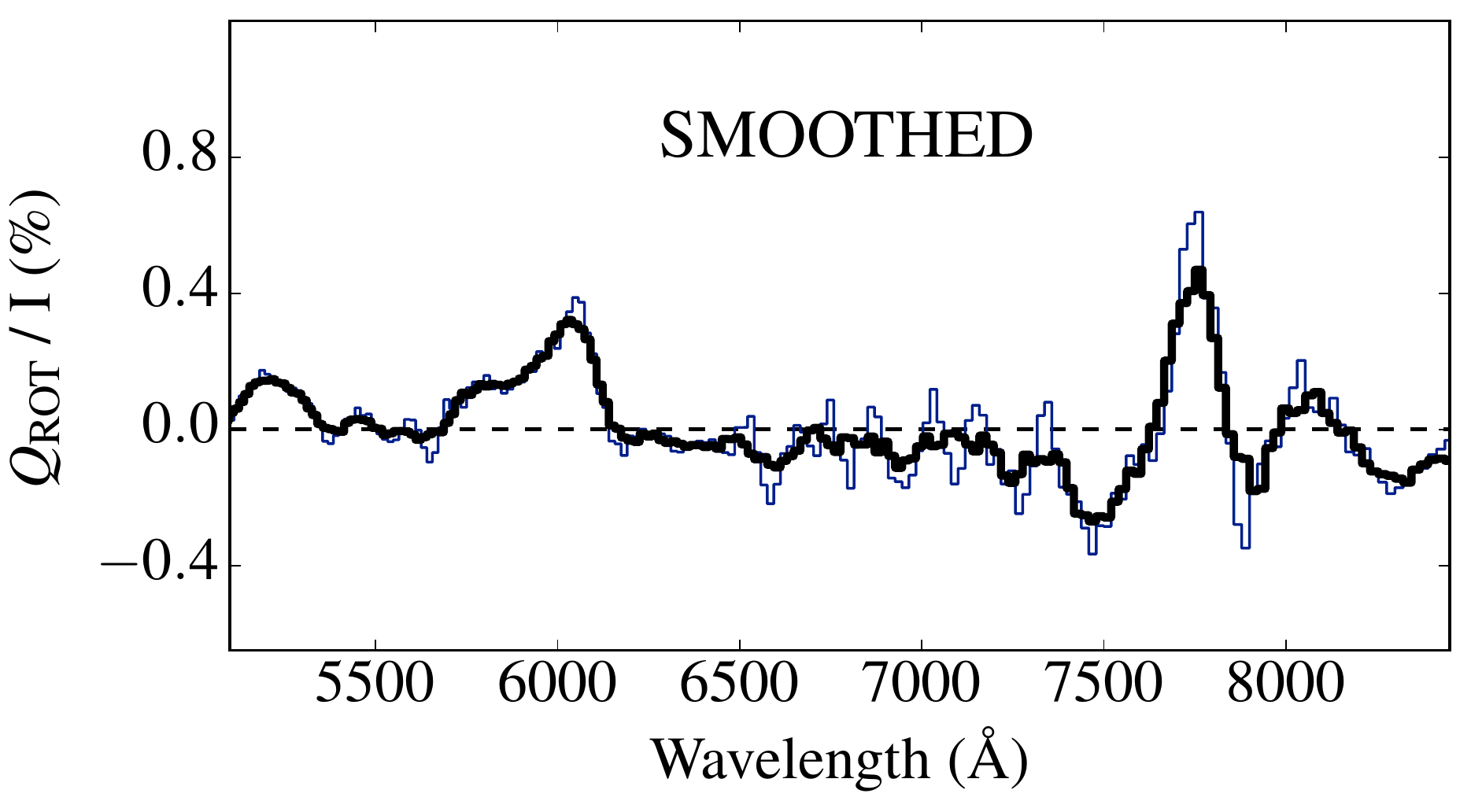}
\includegraphics[width=0.393\textwidth,clip=True,trim=0pt 0pt 0pt -15pt]{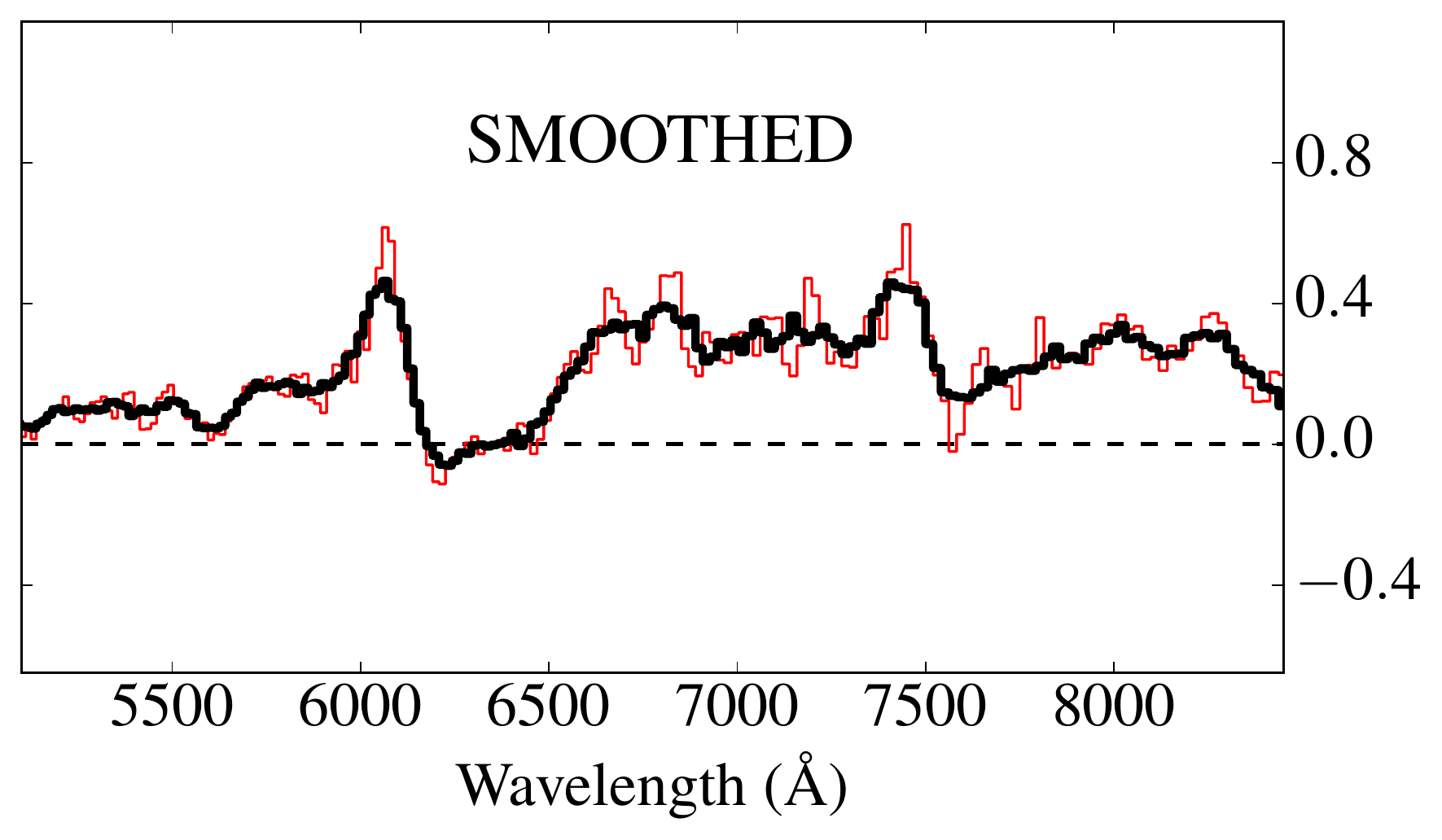}
\caption{\revised{\textit{Upper panels}.} Identification of individual spectral transitions in the 5000$-$8500~\AA{} wavelength region. \revised{Left-hand} panels show flux and polarization spectra for the \subch model viewed from $\bmath{l_1}$, while \revised{right-hand} panels spectra extracted for the \ch model along $\bmath{n_1}$. Spectra are the same as in Fig.~\ref{specmax_subch} and Fig.~\ref{specmax_ddt}. Signals in the 2D \subch model \urot are only due to statistical fluctuations in the simulations and provide estimates of the MC noise levels in all the other polarization spectra (see Section~\ref{noise}). Vertical lines  provide identifications of polarization peaks with spectral transitions. Rest-frame wavelength and blue-shifted velocity of each transition are reported in the corresponding label. Polarization spectra are Savitzky-Golay filtered \revised{using a first-order polynomial and a window} of 3 pixels ($\sim$~50~\AA) for clarity. \revised{\textit{Lower panels}. The same \qrot spectra reported in the upper panels, together with smoothed versions (black) using a Savitzky-Golay filter with a window of 7 pixels ($\sim$~120~\AA). No clear polarization feature is found between 6500 and 7300~\AA.}}
\label{zoom}
\end{center}
\end{figure*}

\begin{figure*}
\begin{center}
\includegraphics[width=0.85\textwidth]{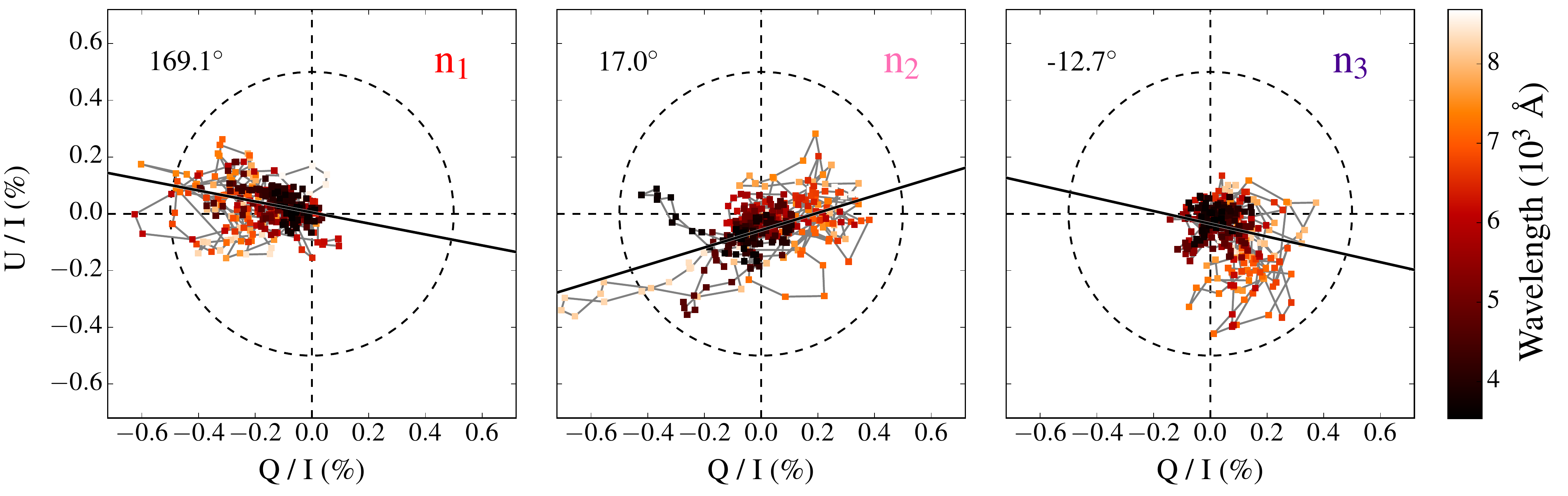}
\caption{$Q/U$ planes for the \ch model at maximum light as viewed from the $\bmath{n_1}$ (left), $\bmath{n_2}$ (middle) and $\bmath{n_3}$ (right) directions. A weighted least-squares fitting to the data is performed for each orientation and the result plotted as a solid black line. The angles determined from the fitting and used to convert $Q$ and $U$ to the polarization spectra of Fig.~\ref{specmax_ddt} are also reported in each panel. Black circles mark polarization levels of 0.5~per~cent. $Q$ and $U$ Stokes parameters are Savitzky-Golay filtered \revised{using a first-order polynomial and a window} of 3 pixels ($\sim$~50~\AA) for clarity. }
\label{quplanes}
\end{center}
\end{figure*}

The bottom panels of Fig.~\ref{specmax_ddt} show polarization spectra \qrot and \urot. By construction (see Section~\ref{noise}), the strongest polarization signal is carried by \qrot for all the viewing angles. However, as the original Stokes parameters are poorly described by a straight line in the $Q/U$ plane (see Fig.~\ref{quplanes}), some residual polarization is also found along \urott. In particular, a simple comparison with the MC noise spectrum of the \subch model (see Fig.~\ref{specmax_subch}) clearly confirms that polarization levels in the \urot spectra of the \ch model can not be solely due to MC noise. The strongest deviations from a single-axis geometry are generally associated with spectral lines, as seen for instance across the Si\,{\sc ii}~$\lambda6355$ feature in the $\bmath{n_3}$ direction.

As in the flux spectrum, the largest line-of-sight variability in the polarization percentage level is found across the Ca~H~and~K and Ca\,{\sc ii} IR triplet profiles. Compared to the other two orientations, the polarization signal along $\bmath{n_2}$ is a factor of $\sim$~3 stronger and reaches peak levels of $\sim$~0.4~per~cent and $\sim$~0.8~per~cent in the Ca~H~and~K and Ca\,{\sc ii} IR triplet, respectively. Some viewing-angle variations are also observed for the Si\,{\sc ii}~$\lambda6355$ line, but now the $\bmath{n_2}$ orientation is associated with the weakest signal (0.3~per~cent) and the $\bmath{n_1}$ observer with the strongest (0.6~per~cent). 

The pseudo-continuum region 6500$-$7500~\AA{} is characterized by polarization levels of about 0.2$-$0.3~per~cent for all the orientations (see also Section~\ref{polevol}). \revised{As shown in Fig.~\ref{zoom} for the $\bmath{n_1}$ orientation, our synthetic spectropolarimetry in this spectral range contains a number of small peaks. Most of these, however, can be ascribed to statistical fluctuations in the simulations since deviations from the mean degree of polarization (\qrot$\sim$~0.3~per~cent) are typically smaller than the average MC noise level in this range ($\sim$~0.2~per~cent, see \urot in the \subch model). This is verified in the lower panels of Fig.~\ref{zoom}, where we plot a smoothed version of the \qrot spectrum.
In the latter, pixel-to-pixel variations due to MC noise are suppressed, while real features are still visible (see e.g. the Si\,{\sc ii}~$\lambda6355$ line). 
We note, however, that the peak around 7500~\AA{} appears real and, as mentioned above for the \subch model, is associated with O\,{\sc i}~$\lambda7774$, Si\,{\sc ii}~$\lambda7849$ and Mg\,{\sc ii}~$\lambda7887$. The nature of this feature will be investigated in the following section.}

%


\subsection{Linking polarization to the ejecta geometry}
\label{geometry}

\begin{figure*}
\begin{center}
\includegraphics[width=0.95\textwidth,clip=True,trim=0pt 0pt -15pt 0pt]{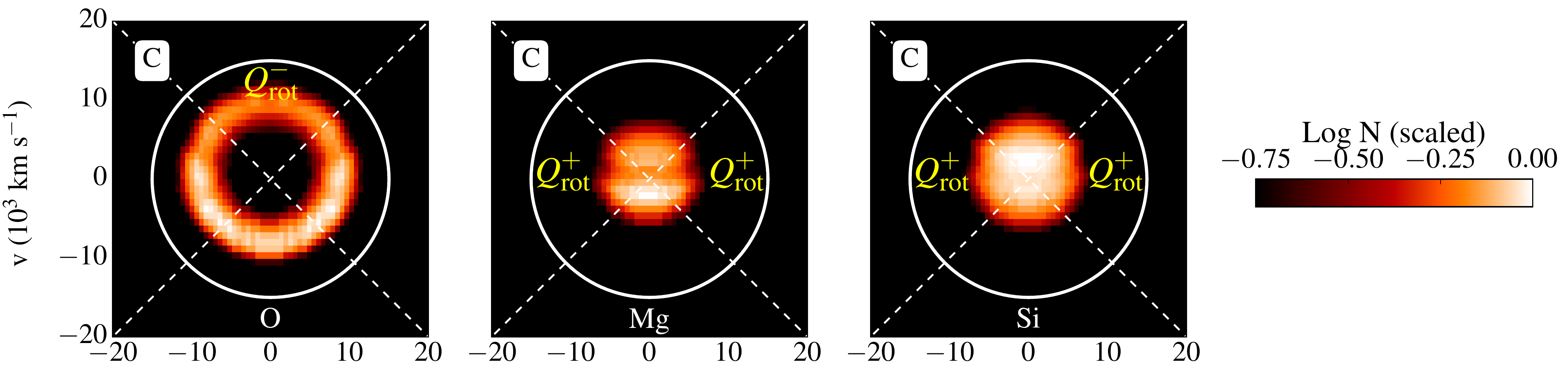}
\includegraphics[width=0.72\textwidth]{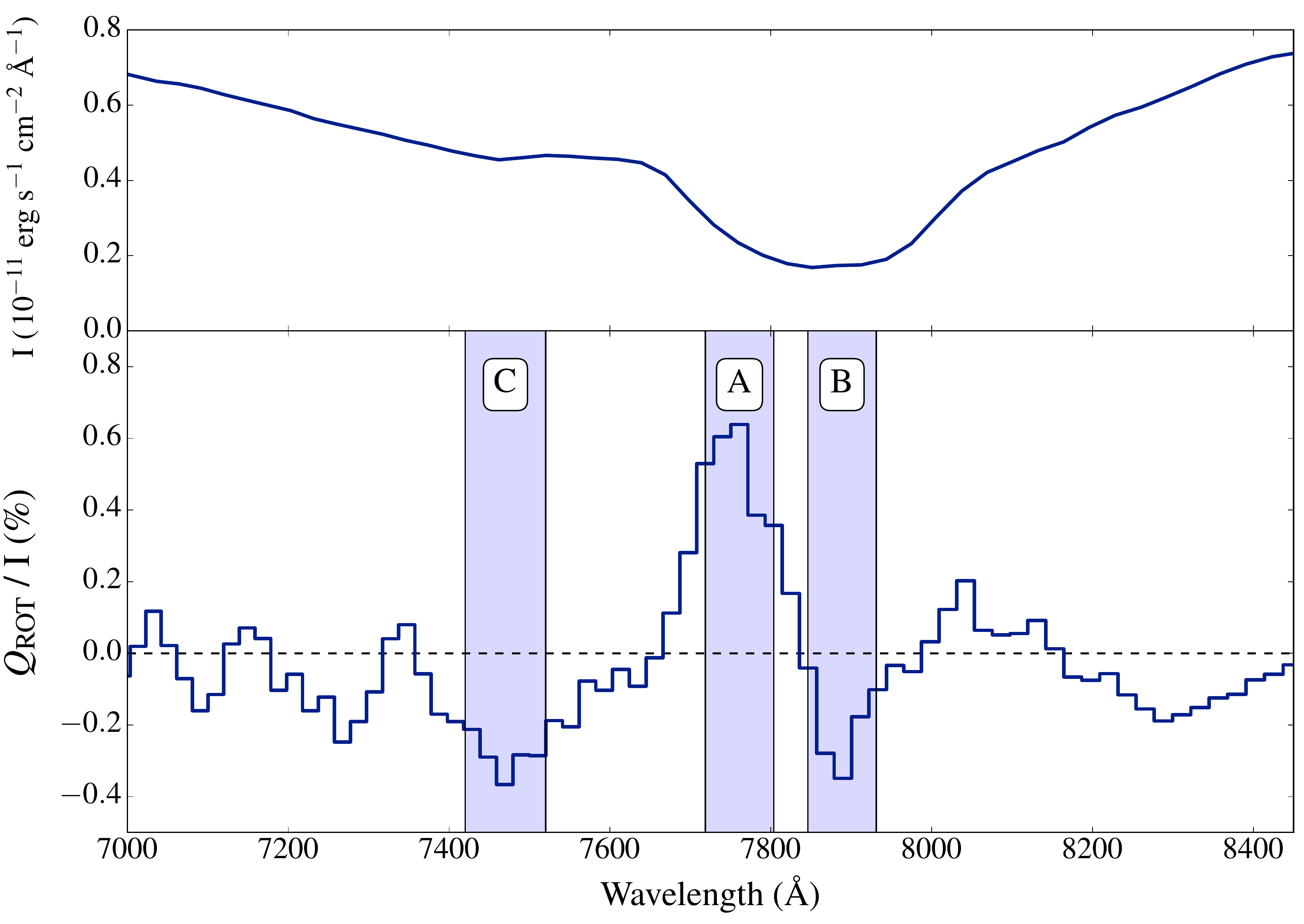}
\includegraphics[width=0.23\textwidth]{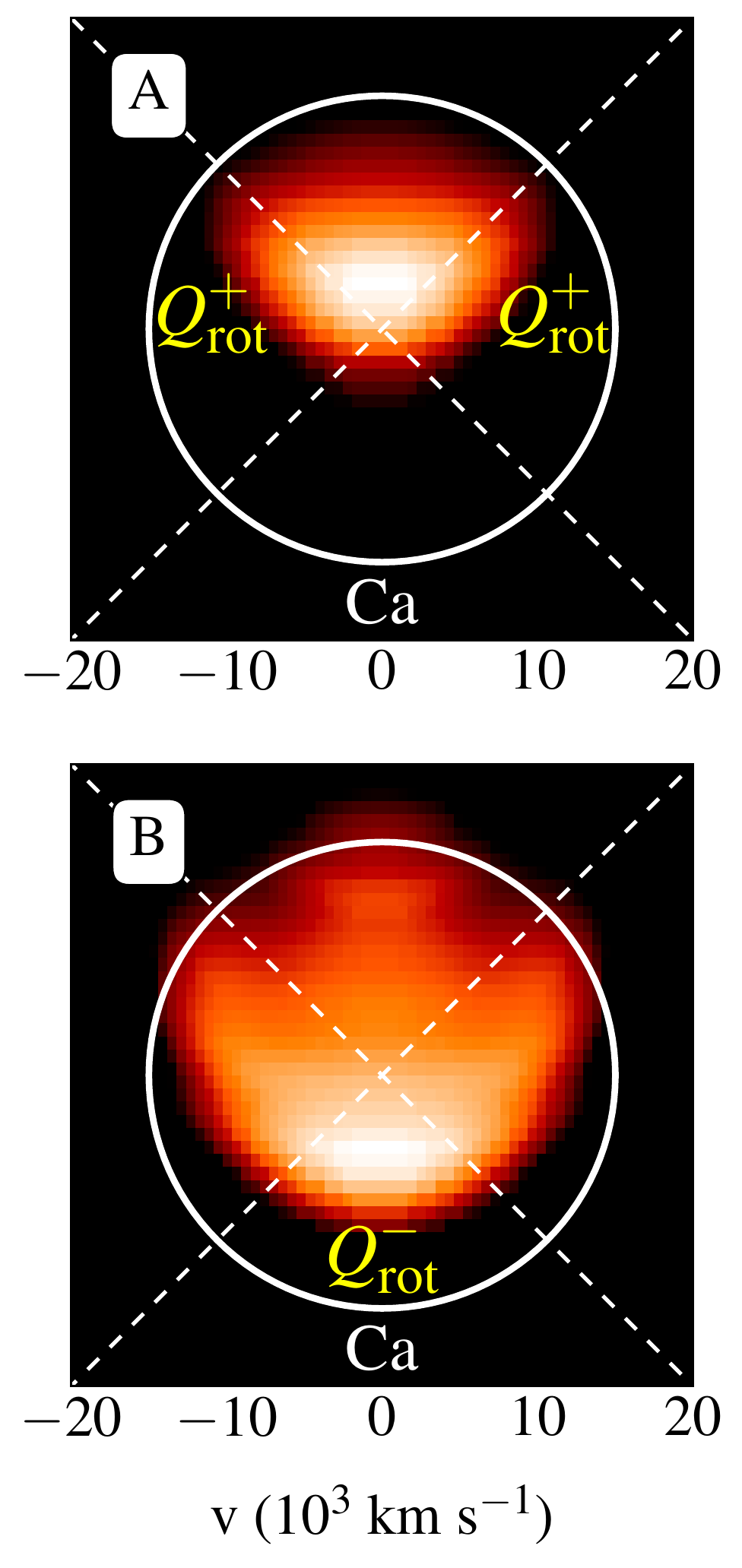}
\caption{\revised{Colour maps of the oxygen, magnesium, silicon (top panels) and calcium (right panels) column densities $N$ calculated through the near-side hemisphere of the ejecta in the \subch $\bmath{l_1}$ direction. As shown by the grey areas in the \qrot spectrum (bottom-left panel), column densities in each map are integrated over different velocity ranges through the ejecta to enclose the polarization peaks in the Ca\,{\sc ii} IR triplet profile (map~A and map~B) and the feature around 7500~\AA{} (map~C). White circles mark a projected velocity of 15\,000~km~s$^{-1}$, within which most of the electron-scattering contributions originate (see Fig.~\ref{lastscatt}). For each map, regions of the ejecta dominating the polarization signal are indicated with $Q_\textrm{rot}^+$ or $Q_\textrm{rot}^-$ if their contribution to \qrot is positive or negative, respectively. The column densities of each map are scaled by the maximum value. Due to the axi-symmetry of the \subch model, maps display a left-right symmetry. }}
\label{geo_subch}
\end{center}
\end{figure*}

\begin{figure*}
\begin{center}
\includegraphics[width=0.97\textwidth,clip=True,trim=-13pt 0pt 0pt 0pt]{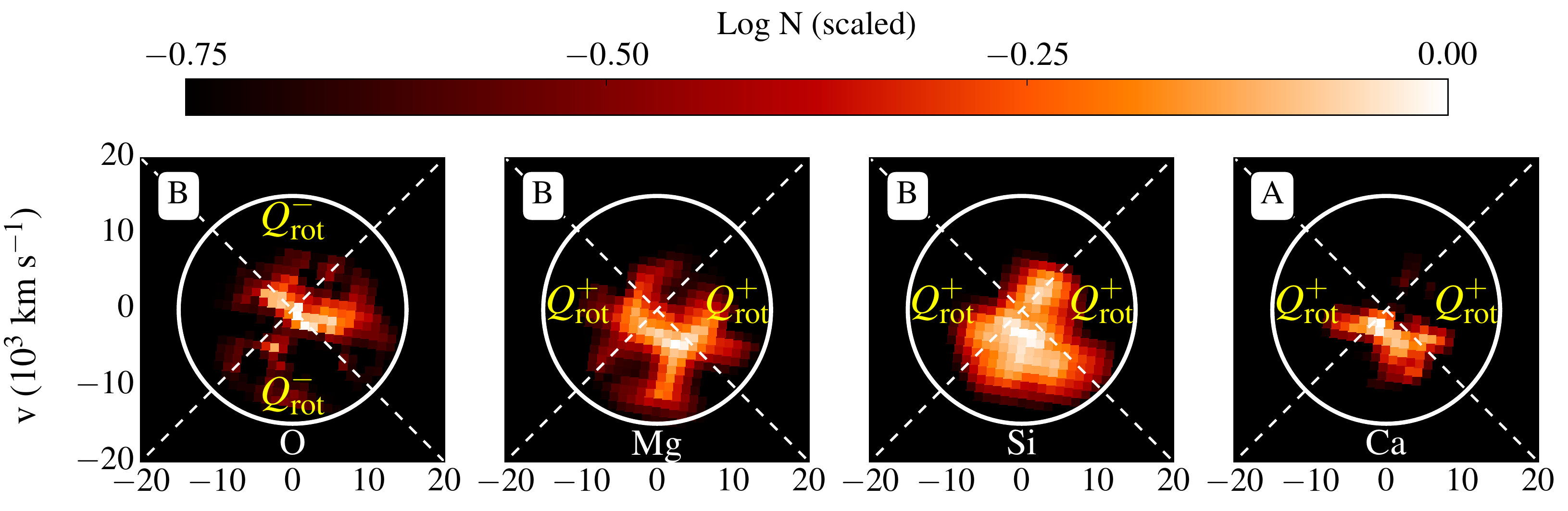}
\includegraphics[width=0.95\textwidth]{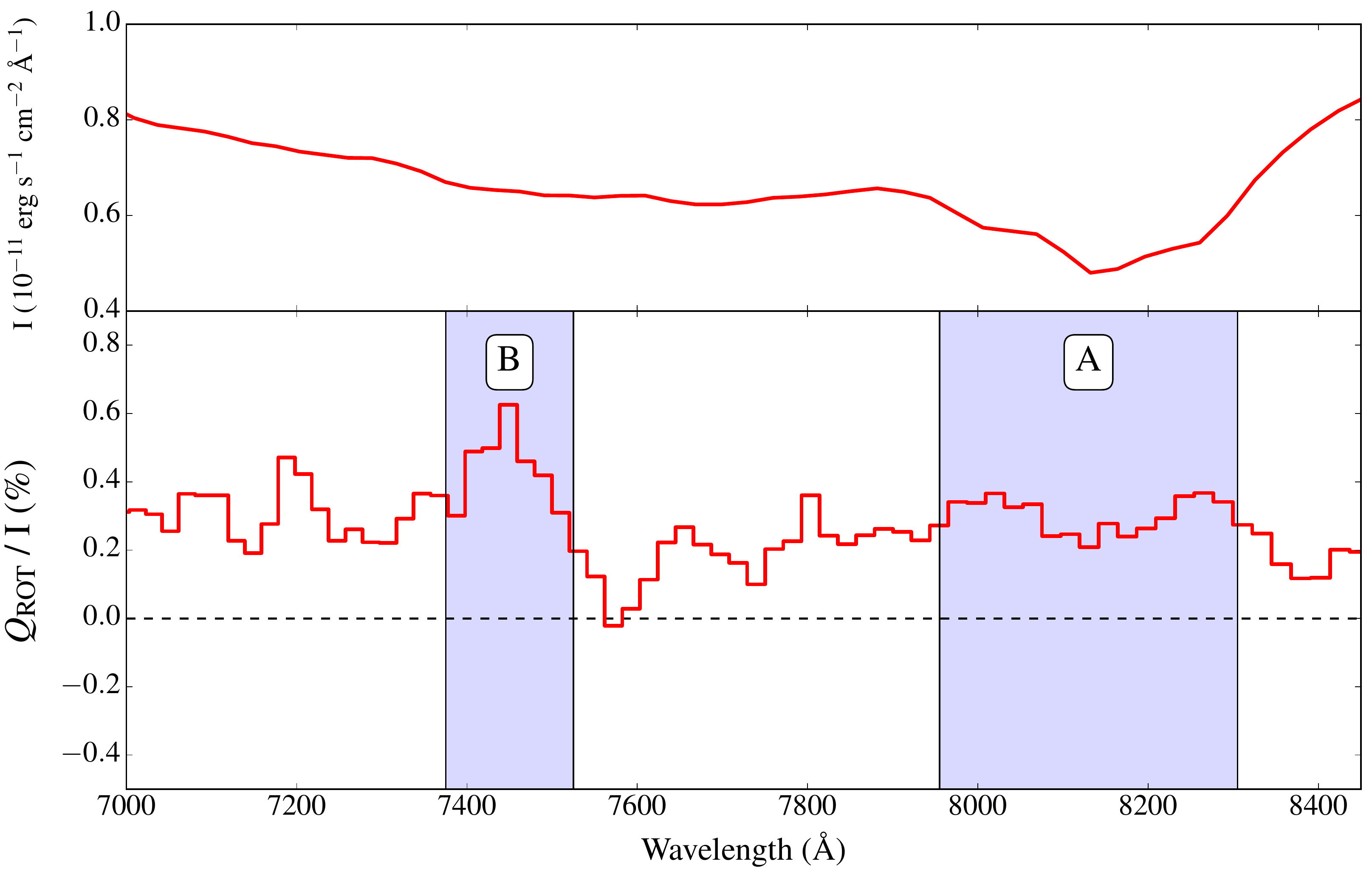}
\caption{\revised{Same as Fig.~\ref{geo_subch}, but for the \ch viewed along the $\bmath{n_1}$ direction. Colour maps have been rotated of $\alpha_\textrm{rot}/2=84.5^\circ$ (see left panel of Fig.~\ref{quplanes}) in the clockwise direction to account for the transformation of $Q$ and $U$ to \qrot and \urot (see Section~\ref{noise}).}}
\label{geo_ddt}
\end{center}
\end{figure*}

\revised{In this section, we will examine more closely how asymmetries in the element distribution are linked to individual polarization features. In particular, we focus on two examples. First, we investigate the link between the calcium distribution in the ejecta and the polarization signatures detected across the Ca\,{\sc ii} IR triplet profile. Second, we look at the oxygen, magnesium and silicon morphologies to understand which of these elements are responsible for the polarization features observed around 7500~\AA{} (see Section~\ref{polmax}). We frame our discussion in terms of the physical picture in which polarization variations across spectral lines can be understood as resulting from partial covering of the electron-scattering photosphere caused by the asymmetric distributions of elements in the ejecta \citep[see e.g.][]{kasen2003}.}

\subsubsection{2D \subch model}
\label{geometry_subch}

\begin{figure*}
\begin{center}
\includegraphics[width=0.98\textwidth]{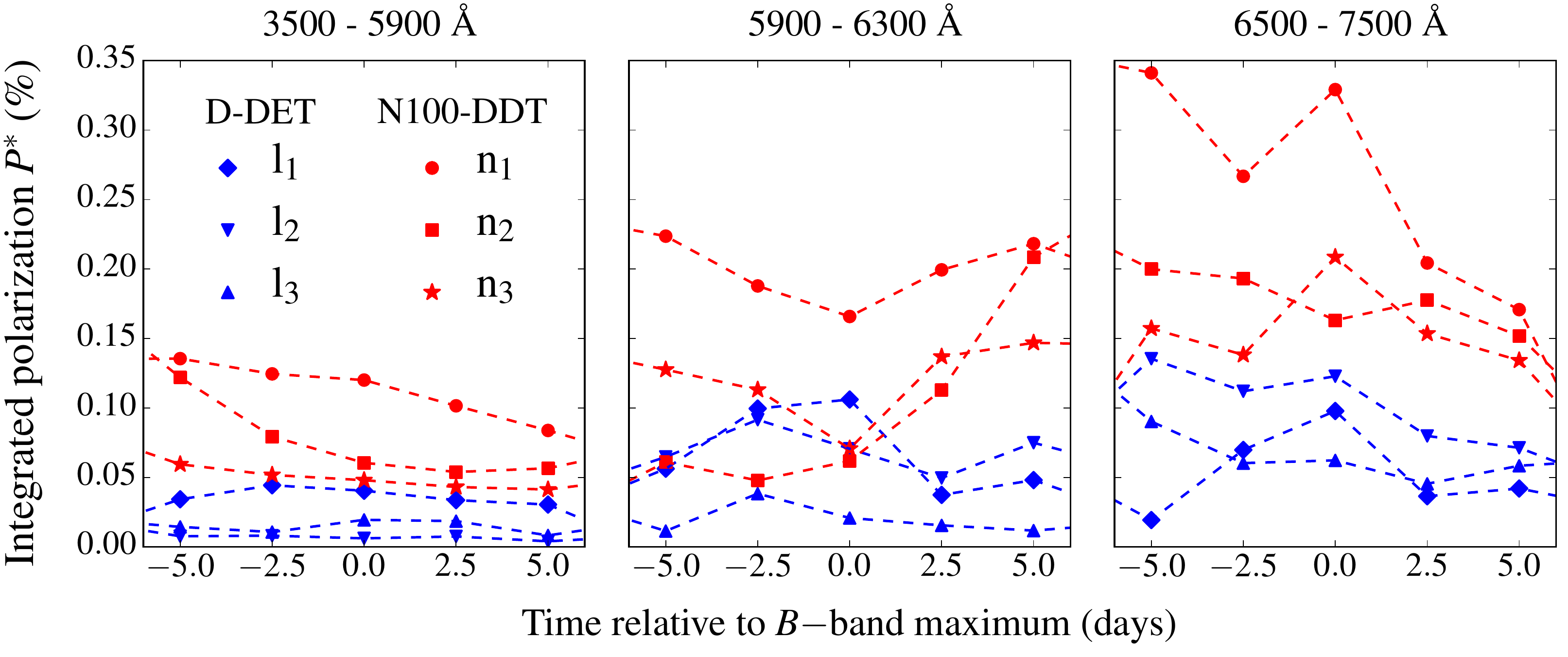}
\caption{Polarization light curves between $-$5 and +5~d relative to $B-$band maximum light calculated in three different wavelength ranges: 3500$-$5900~\AA{} (left-hand panel,``blue'' range), 5900$-$6300~\AA{} (middle panel, around the Si\,{\sc ii} $\lambda6355$ line) and 6500$-$7500~\AA{} (right-hand panel, pseudo-continuum range). MC noise error bars are $\sim$~0.001, 0.013 and 0.029\revised{~per~cent} in the left-hand, middle and right-hand panel, respectively.}
\label{vlcpol}
\end{center}
\end{figure*}

\revised{As shown in Fig.~\ref{geo_subch}, the polarization signature extracted across the Ca\,{\sc ii} IR triplet in the $\bmath{l_1}$ direction is characterized by a transition from positive to negative \qrot levels at around 7830~\AA{} (corresponding to velocities of about $-$23\,500~km~s$^{-1}$). This transition can be understood by inspecting the calcium distribution in the \subch ejecta more closely. The colour maps in the right panels of Fig.~\ref{geo_subch} present column densities ($N$) for Ca calculated through the near-side hemisphere of the ejecta along the  \subch $\bmath{l_1}$ direction. 
We first notice that the distribution of calcium at very high velocities ($-$26\,000~km~s$^{-1}$, map A) preferentially covers the top side of the ejecta as viewed from the $\bmath{l_1}$ orientation. Most of the electron-scattered contributions from this side of the ejecta -- mainly polarized in the negative \qrot direction -- are thus absorbed by calcium and do not reach the observer. As a result, positive contributions from the left and right side of the ejecta dominate the signal and produce a polarization peak with \qrot$>$~0 in the Ca\,{\sc ii} IR triplet profile. If we now move deeper into the ejecta ($-$21\,500~km~s$^{-1}$, map B), we notice that here the top, left and right sides are preferentially obscured by calcium. Hence, packets are more likely to escape from the bottom side of the ejecta and bias the polarization signal towards negative values (\qrot$<$~0) in the Ca\,{\sc ii} IR profile.}

\revised{As pointed out in Section~\ref{subchmax}, the spectrum extracted along the $\bmath{l_1}$ orientation is characterized by a clear polarization feature around 7500~\AA. This feature is polarized in the negative \qrot direction, meaning that its polarization angle is 90$^\circ$ rotated from most of the other features in the spectrum (e.g. the Si\,{\sc ii}~$\lambda6355$ line, see Fig.~\ref{zoom}). Three spectral transitions can be responsible for the variation in polarization across this spectral region: O\,{\sc i}~$\lambda7774$, Si\,{\sc ii}~$\lambda7849$ or Mg\,{\sc ii}~$\lambda7887$. As shown in the upper panels of Fig.~\ref{geo_subch}, two distinct behaviours are seen in the distributions of oxygen, magnesium and silicon: the asymmetric distributions of magnesium and silicon will cause positive contributions from the left and right sides of the ejecta to dominate the polarization signal\footnote{Note that this is consistent with the positive signals detected across the silicon lines, see Fig.~\ref{zoom}.}. On the other hand, the distribution of oxygen will favour negative contributions (from the top side of the ejecta) dominating the signal. Therefore -- although all the three aforementioned transitions are expected to affect the degree of polarization to some extent -- we suggest that the polarization signal extracted around 7500~\AA{} in the $\bmath{l_1}$ direction is dominated by O\,{\sc i}~$\lambda7774$. This identification explains why we see a 90$^\circ$ rotation between the feature around 7500~\AA{} and the silicon lines (e.g. Si\,{\sc ii}~$\lambda5972$ and Si\,{\sc ii}~$\lambda6355$, see Fig.~\ref{zoom}).}

\subsubsection{3D \ch model}
\label{geometry_ddt}

\revised{As shown in the bottom panels of Fig.~\ref{geo_ddt}, positive \qrot polarization levels are detected across the Ca\,{\sc ii} IR profile for the \ch $\bmath{n_1}$ direction. This is readily understood by looking at the distribution of calcium in the ejecta of the \ch model (map~A, top right panel of Fig.~\ref{geo_ddt}). When viewed from the $\bmath{n_1}$ observer orientation, the asymmetric distribution of calcium offers higher opacities to packets escaping towards the observer from the bottom side of the ejecta. As a result, the polarization signal across the Ca\,{\sc ii} IR profile is dominated by contributions coming from the left and right sides and thus biased towards positive values. }

\revised{As mentioned in Section~\ref{chmax}, the polarization spectrum extracted for the \ch model is characterized by a signature in the wavelength region occupied by the O\,{\sc i}~$\lambda7774$, Si\,{\sc ii}~$\lambda7849$ and Mg\,{\sc ii}~$\lambda7887$ lines. Specifically, this feature is polarized in the positive \qrot direction when viewed from the $\bmath{n_1}$ direction. As shown in the top panels of Fig.~\ref{geo_ddt} (maps~B), magnesium and silicon column densities extracted along $\bmath{n_1}$ are higher on the bottom side of the ejecta, thus favouring positive contributions to dominate the \qrot polarization signal. In contrast, oxygen preferentially absorbs packets from the right side of the ejecta and thus biases the \qrot polarization signal towards negative values. Based on this analysis, we therefore suggest that -- unlike what is seen for the \subch model -- the polarization signal across the 7500~\AA{} feature is dominated by Si\,{\sc ii}~$\lambda7849$ and Mg\,{\sc ii}~$\lambda7887$. Although the contribution of O\,{\sc i}~$\lambda7774$ appears not to be dominant in this case -- i.e. around maximum light and towards the $\bmath{n_1}$ observer orientation -- we note that the same might not be true for different directions and at different epochs (see Section~\ref{poloxygen}).  }

\subsection{Polarization light curves}
\label{polevol}

In this section, we investigate the temporal evolution of polarization for the \subch and \ch models. Following \citet{bulla2015}, we calculate intensity ($I^*$) and polarization (\qrott$^*$ and \urott$^*$) light curves by integrating the flux and polarization spectra over selected wavelength ranges [$\lambda_1,\lambda_2$]:
\begin{equation}
\begin{bmatrix} I^*(t) \\ \qrott^*(t) \\ \urott^*(t) \end{bmatrix} = \bigintsss_{\lambda_1}^{\lambda_2} \begin{bmatrix} I(\lambda,t) \\ \qrott(\lambda,t) \\ \urott(\lambda,t) \end{bmatrix}  ~\mathrm{d}\lambda~.
\end{equation}
Specifically, we focus our attention on three spectral regions: from 3500 to 5900~\AA{} (``blue'' range), from 5900 to 6300~\AA{} (Si\,{\sc ii} $\lambda6355$ range) and from 6500 to 7500~\AA{} (pseudo-continuum range, see above). Given that MC noise levels integrated in these three wavelength ranges are low (0.001, 0.013 and 0.029~per~cent, respectively), polarization bias (see Section~\ref{noise}) is less an issue here. Therefore, we restrict our discussion to the time evolution of the integrated polarization percentage $P^*$ defined as
\begin{equation}
P^*(t) = \frac{\sqrt{\qrott^*(t)^2+\urott^*(t)^2}}{I^*(t)}~.
\end{equation}
Fig.~\ref{vlcpol} reports $P^*$ around $B-$band maximum light for the \subch and \ch models in the three selected ranges.

First, we notice that polarization levels for the \subch model are generally lower in all ranges and at all epochs compared to those found for the \ch model. Second, we find pseudo-continuum polarization levels for both models (right panel of Fig.~\ref{vlcpol}) to vary in the range between 0.05 and 0.3~per~cent, in good agreement with spectropolarimetric observations of SNe Ia \citep{wang2008}. Finally, both models are characterized by an increase in the degree of polarization with wavelength, a behaviour that can not be ascribed solely to MC noise growing from blue to red regions of the spectra (see above). As proposed by \citet{wang1997} and observed in several SNe~Ia, this rise in polarization towards the red occurs as a result of the decreasing contribution of lines to the opacity\footnote{Line scattering in SNe is usually assumed to depolarize the incoming radiation. Although resonance line scattering can lead to a polarization contribution \citep{hamilton1947}, the latter is typically lower than that from electron scattering and thus usually considered as a second-order effect \citep{jeffery1989,jeffery1991}.}. In our simulations the fraction of escaping packets that had a line interaction as their last event are $\sim$~75, 50 and 30~per~cent in the 3500$-$5900~\AA, 5900$-$6300~\AA{} and 6500$-$7500~\AA{} ranges, respectively.

The low MC noise levels in the integrated quantities defined above allows us to study the polarization evolution around maximum. In all the spectral ranges, we find polarization levels for the \subch model to reach their peak between 0 -- 5~d before $B-$band maximum light, and to decrease thereafter. This behaviour is consistent with what is seen in SNe~Ia and usually ascribed to the photosphere transitioning from asymmetric outer ejecta to more spherical inner regions \citep{wang2008}. Qualitatively similar results are found for the \ch model, although $P^*$ across the Si\,{\sc ii}~$\lambda6355$ profile continues to increase after maximum light, in this case (middle panel in Fig.~\ref{vlcpol}).


\section{Comparison with observations}
\label{compobs}

Here we compare our model spectra with observational data of normal SNe~Ia. In Section~\ref{compsn} we present comparisons between full polarization spectra predicted by our models and those observed in normal SNe~Ia. \revised{In Section~\ref{polsilicon}, Section~\ref{poloxygen} and Section~\ref{polcalcium} we then focus our attention on the polarization signals across the Si\,{\sc ii}~$\lambda6355$ line, the O\,{\sc i}~/~Si\,{\sc ii}~/~Mg\,{\sc ii} feature around 7500~\AA{} and the Ca\,{\sc ii} IR profile, respectively}. \revised{Polarization values $P$ are used throughout this section, as has been common practice in previous studies including those that present the observational work to which we will compare (see also Section~\ref{noise}).}

\subsection{Comparison with SN Ia polarization spectra}
\label{compsn}

In the following we compare flux and polarization spectra predicted for the \subch and \ch model with data of three well-studied SNe Ia: SN~2012fr, SN~2001el and SN~2004dt. These supernovae are selected as they have been observed in polarization at multiple epochs and with high spectral signal-to-noise ratio. In addition, the polarization range spanned by these objects encompasses the distribution observed for normal SNe~Ia, going from the rather low levels observed for SN~2012fr to a high degree of polarization in SN~2004dt. Unless otherwise stated, we present synthetic spectra for the two orientations (one for the \subch and one for the \ch model) that provide best matches with observed flux and polarization spectra. We note, however, that spectra in better agreement with observations might be found for viewing angles not investigated in this study.

\begin{figure*}
\begin{center}
\includegraphics[width=0.94\textwidth]{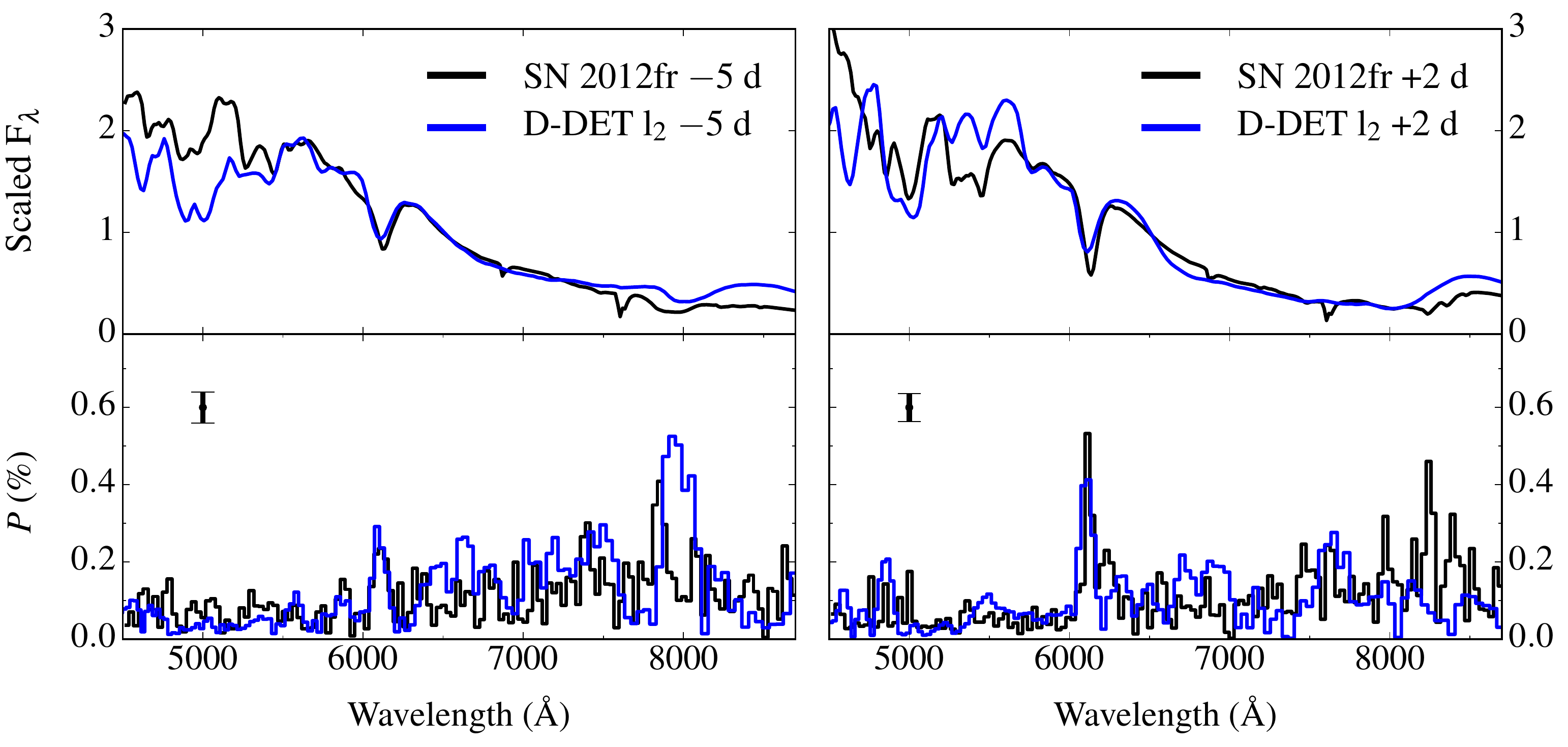}
\includegraphics[width=0.94\textwidth]{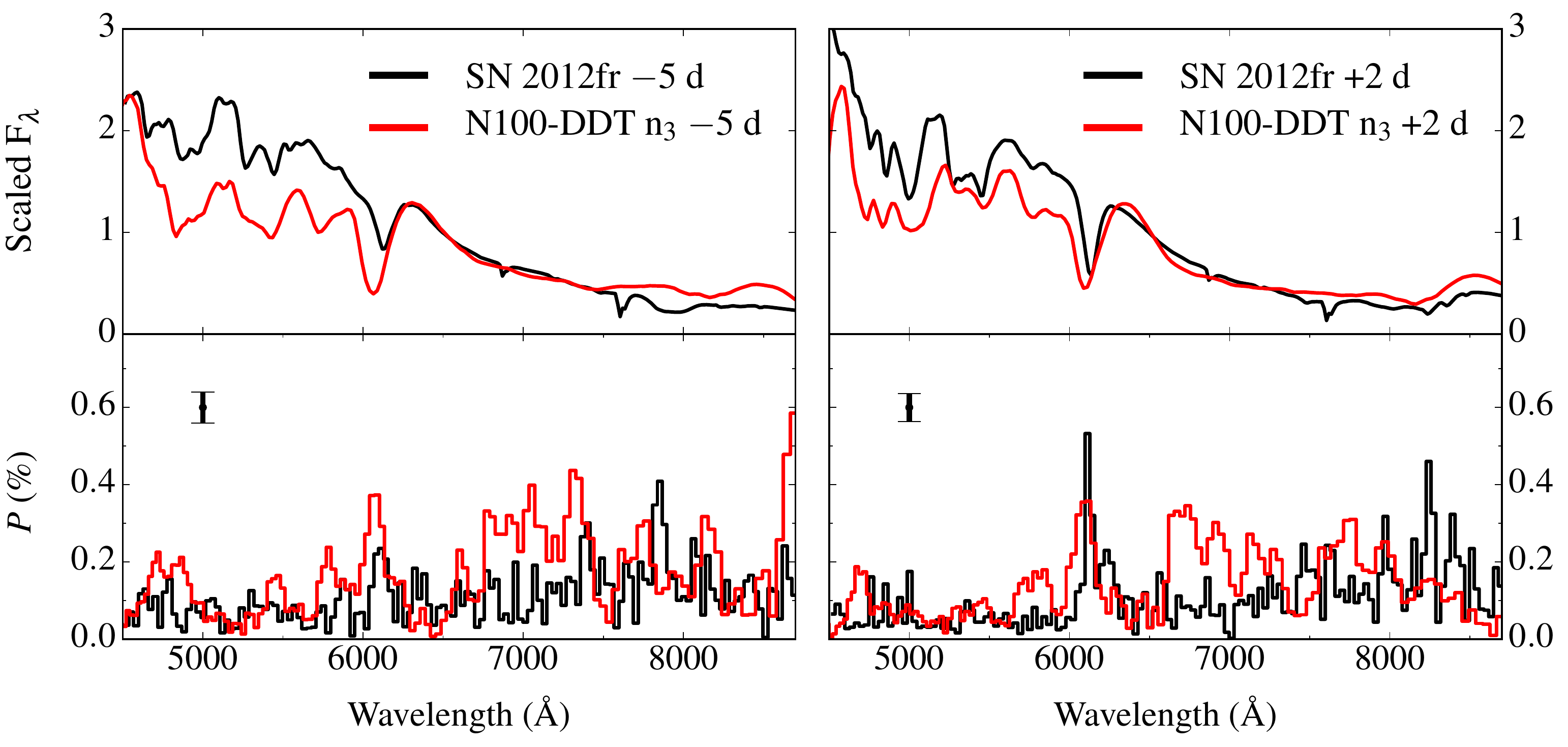}
\caption{Flux and polarization spectra predicted for the \subch (upper panels, blue lines) and \ch (lower panels, red lines) models viewed from the $\bmath{l_2}=(1,0,0)$ and $\bmath{n_3}=(-\nicefrac{1}{2},\nicefrac{1}{2},\nicefrac{1}{\sqrt{2}})$ orientations, respectively. Spectra are reported at $-$5 (left-hand panels) and +2 (right-hand panels) d relative to $B-$band maximum light. For comparison, black lines show observed spectra of SN~2012fr at the same epochs after correction for the interstellar polarization component \citep[$\sim$~0.24~per~cent;][]{maund2013}. Black bars mark the averaged errors in the observed polarization spectra. \revised{Both synthetic and observed flux spectra have been normalized to their values at 6500~\AA{} for illustrative purposes.}}
\label{2012fr}
\end{center}
\end{figure*}

\begin{figure*}
\begin{center}
\includegraphics[width=0.95\textwidth]{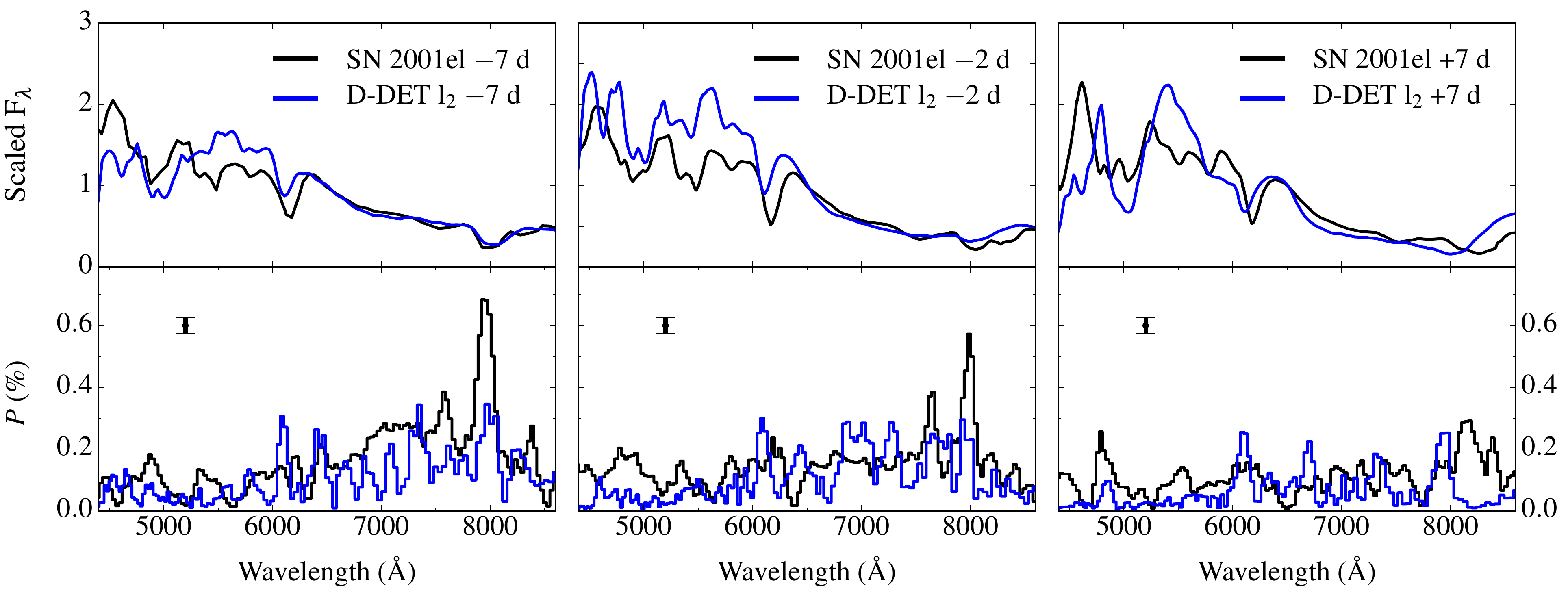}
\includegraphics[width=0.95\textwidth]{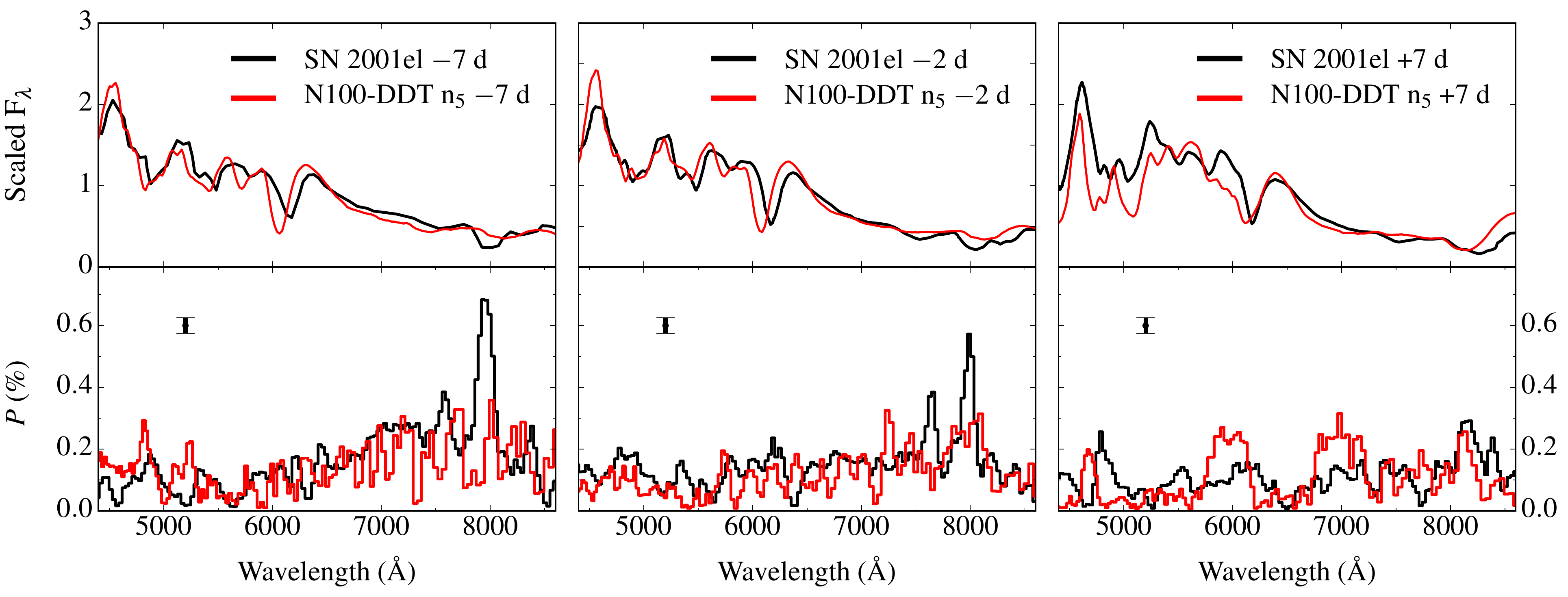}
\caption{Flux and polarization spectra predicted for the \subch (upper panels, blue lines) and \ch (lower panels, red lines) models viewed from the $\bmath{l_2}=(1,0,0)$ and $\bmath{n_5}=(\nicefrac{1}{\sqrt{2}},-\nicefrac{1}{\sqrt{2}},0)$ directions, respectively. Spectra are reported at $-$7 (left-hand panels), $-$2 (middle panels) and +7 (right-hand panels) d relative to $B-$band maximum light. For comparison, black lines show observed spectra of SN~2001el at the same epochs after correction for the interstellar polarization component \citep[$\sim$~0.6~per~cent;][]{wang2003}. Black bars mark the averaged errors in the observed polarization spectra. \revised{Both synthetic and observed flux spectra have been normalized to their values at 6500~\AA{} for illustrative purposes.}}
\label{2001el}
\end{center}
\end{figure*}

\begin{figure*}
\begin{center}
\includegraphics[width=0.98\textwidth]{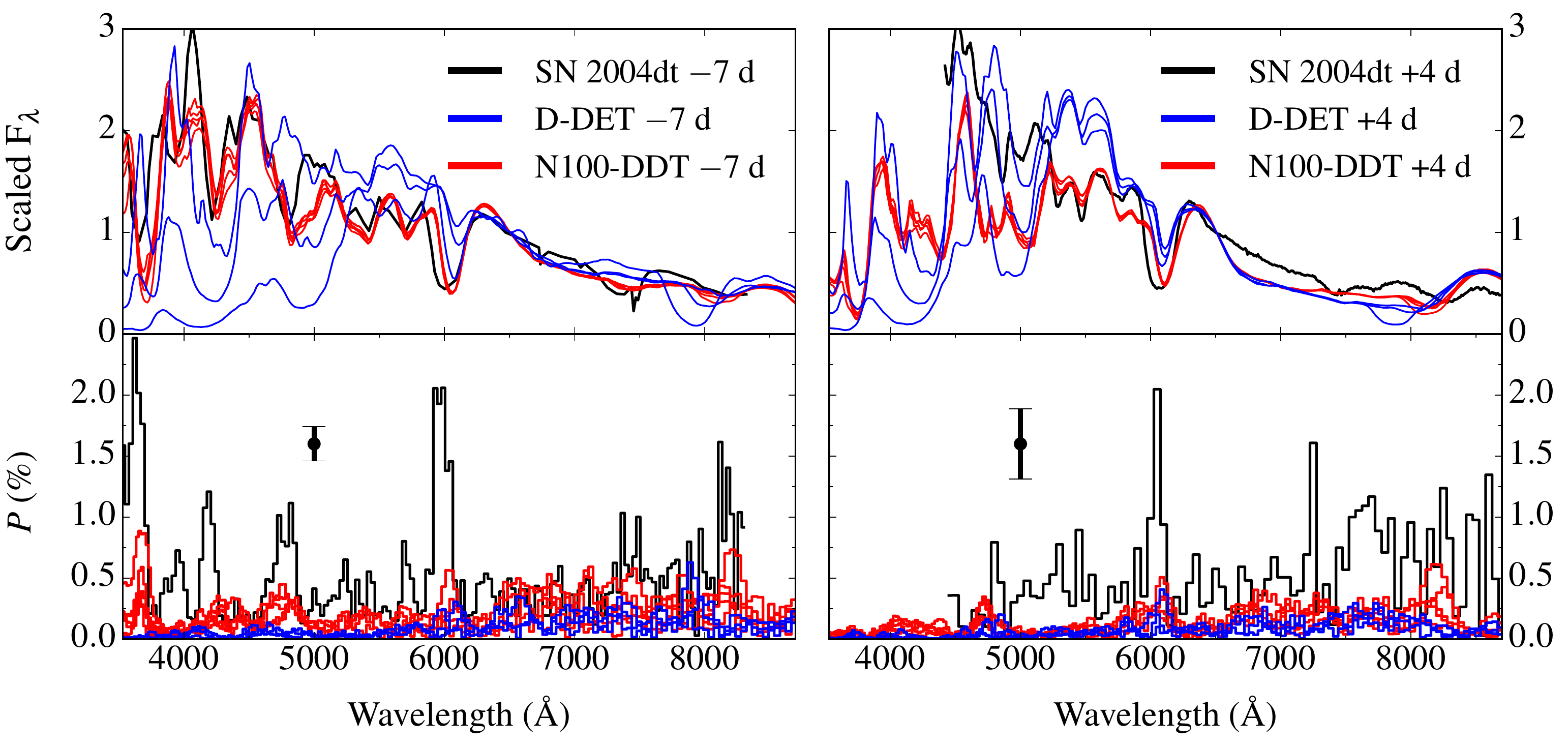}
\caption{Flux (top) and polarization (bottom) spectra predicted for the \subch (three blue lines) and the \ch (five red lines) models for all the high signal-to-noise calculation orientations investigated in this paper. Spectra are reported at $-$7 (left-hand panels) and +4 (right-hand panels) d relative to $B-$band maximum. For comparison, black lines show observed spectra of SN~2004dt at the same epochs after correction for the interstellar polarization component \citep[$\sim$~0.28~per~cent;][]{leonard2005,wang2006}. Black bars mark the averaged errors in the observed polarization spectra. \revised{Both synthetic and observed flux spectra have been normalized to their values at 6500~\AA{} for illustrative purposes.}}
\label{2004dt}
\end{center}
\end{figure*}

\subsubsection{SN~2012fr}
\label{sec2012fr}

SN~2012fr was observed in polarization using the Focal Reducer and low dispersion Spectrograph (FORS1) at the ESO Very Large Telescope (VLT). Polarization levels were found to be generally very small ($\sim$~0.1~per~cent in the pseudo-continuum range), although higher signals of $\sim$~0.5$-$0.6~per~cent were detected across the Si\,{\sc ii}~$\lambda6355$ and Ca\,{\sc ii} IR triplet features \citep{maund2013}. In particular, a clear transition from high- (pre-maximum) to low- (post-maximum) velocity components was observed in the calcium profile. In Fig.~\ref{2012fr}, we show spectropolarimetric data of SN~2012fr at $-$5 and +2~d relative to $B-$band maximum together with synthetic spectra extracted for the \subch and \ch models at the same epochs. The best match between predicted and observed spectra is found for the \subch $\bmath{l_2}$ (upper panels) and \ch $\bmath{n_3}$ (lower panels) orientations.

There is a remarkably good match between the \subch model and SN~2012fr at both epochs. Flux spectra of SN~2012fr are reproduced rather well by this model, both in terms of spectral shapes and velocities. The agreement is particularly good at 5~d before maximum, although we note two important discrepancies: the flux spectrum is too red and the velocity of Ca\,{\sc ii} IR triplet too low in our simulations. As discussed by \citet{kromer2010}, the red colour can be attributed to the composition of the burning yields in the outer shell. Specifically, the moderate amount of IGEs produced in the helium shell (see Fig.~\ref{ejecta}) leads to strong line blanketing in the blue spectral regions and flux redistribution to redder wavelengths (see also discussion in Section~\ref{subchmax}). The discrepancy in the velocity of the Ca\,{\sc ii} IR triplet is directly related to the specific distribution of calcium in the outer regions of the \subch ejecta. Velocities of $\sim$~17\,000~km~s$^{-1}$ measured in the synthetic profile are consistent with the calcium opacity predicted along the $\bmath{l_2}$ orientation (see Fig.~\ref{ejecta}), but too small to reproduce those observed in the high-velocity features of SN~2012fr ($\sim$~20\,000~km~s$^{-1}$\revised{, but see also Section~\ref{polcalcium}}).


The agreement between the \subch model and observations is even more striking in terms of polarization. The match in the pseudo-continuum range (6500$-$7500~\AA) and in the bluer regions of the spectra ($\lesssim$~6000~\AA) is remarkable both pre- and post-maximum. In addition, the degrees of polarization observed across the Si\,{\sc ii}~$\lambda6355$ line are nearly identical to those predicted by the \subch model at $-$5 ($\sim$~0.3~per~cent) and +2 ($\sim$~0.4~per~cent) d relative to maximum. In contrast, we notice a poor match between model and data for the only other feature clearly detected in SN~2012fr, the Ca\,{\sc ii} IR triplet. While the calcium profile in SN~2012fr is characterized by a transition from a high-velocity component to three distinct low-velocity components, our model predicts a fairly strong and broad feature 5~d before maximum that almost disappears a week after. We note, however, that since the model fails to provide a convincing match to the Ca IR triplet in the flux spectrum (see above), it is not surprising that the match in polarization across this feature is also relatively poor.

As shown in the bottom panels of Fig.~\ref{2012fr}, the degree of polarization predicted for the \ch model is broadly consistent with the levels observed in SN~2012fr. The polarization signals in the blue part of the spectrum and across the Si\,{\sc ii}~$\lambda6355$ profile agree reasonably well with the data. However, we see larger discrepancies compared to those found in the \subch case. First, the model predicts velocities along $\bmath{n_3}$ that are typically higher than those observed (e.g. 30~per~cent larger for the Si\,{\sc ii}~$\lambda6355$ profile 5~d before maximum). Second, velocities for the calcium IR triplet ($\sim$~13\,000~km~s$^{-1}$) are definitely too low to account for the high-velocity component observed in SN~2012fr (\revised{see also Section~\ref{polcalcium}}). Finally, the 6500$-$7500~\AA{} pseudo-continuum range in the \ch model spectrum \revised{is too highly polarized ($\sim$~0.3~per~cent) at both epochs to match the small signal of $\sim$~0.1~per~cent detected in SN~2012fr.}

\subsubsection{SN~2001el}
\label{sec2001el}

Spectropolarimetric observations of SN~2001el were triggered with FORS1 at ESO VLT. A polarization signal of $\sim$~0.2$-$0.3~per~cent -- typical for the bulk of normal SN~Ia events -- was detected in the pseudo-continuum range \citep{wang2003}. Two distinct components were clearly visible in the Ca\,{\sc ii} IR triplet: a significantly polarized ($\sim$~0.7~per~cent) high-velocity component and a distinct low-velocity component polarized at $\sim$~0.2$-$0.3~per~cent level. In Fig.~\ref{2001el}, we compare observations of SN~2001el at $-$7, $-$2 and +7~d relative to $B-$band maximum with synthetic spectra extracted for the \subch (upper panels) and \ch (lower panels) model at the same epochs. Specifically, we report the best matches found in each case, which are for the \subch $\bmath{l_2}$ and the \ch $\bmath{n_5}$ orientations.

Spectra extracted for the \subch model along the equatorial $\bmath{l_2}$ orientation are broadly consistent with those observed in SN~2001el. Polarization levels extracted in the pseudo-continuum region and across the Si\,{\sc ii}~$\lambda6355$ profile are in good agreement with data, although velocities in the synthetic spectra are slightly larger than those observed. The most striking difference, however, concerns the Ca\,{\sc ii} IR triplet region. The calcium distribution in the outer shell of the \subch model (see Fig.~\ref{ejecta}) is able to reproduce the high-velocity component seen in SN~2001el, but is not sufficiently asymmetric to explain the relatively high polarization level of 0.6$-$0.7~per~cent observed. In this respect, it is worth noting that such strong polarization signals are predicted -- at least around maximum -- for the $\bmath{l_1}$ orientation of the \subch model (see for instance Fig.~\ref{specmax_subch}). However, this orientation is characterized by (i) spectra that are much too red and by (ii) a calcium line that is too strong and too fast to match the observations (\revised{see discussion in Section~\ref{polcalcium}}).

Spectra extracted for the \ch model along $\bmath{n_5}$ agree reasonably well with data of SN~2001el, although we again notice systematically larger velocities in the model. The match is particularly good before maximum light, with the predicted spectral color \citep[see also][]{roepke2012} and degrees of polarization found to be consistent with observations. As in the \subch model, however, the region around the Ca\,{\sc ii} IR triplet is poorly reproduced by the \ch model. Here, the polarization signal is a factor of $\sim$~2 smaller than that observed. Furthermore, the velocity is also too slow to reproduce the high-velocity component seen in SN~2001el before maximum.

\subsubsection{SN~2004dt}
\label{sec2004dt}

\begin{figure*}
\begin{center}
\includegraphics[width=0.98\textwidth]{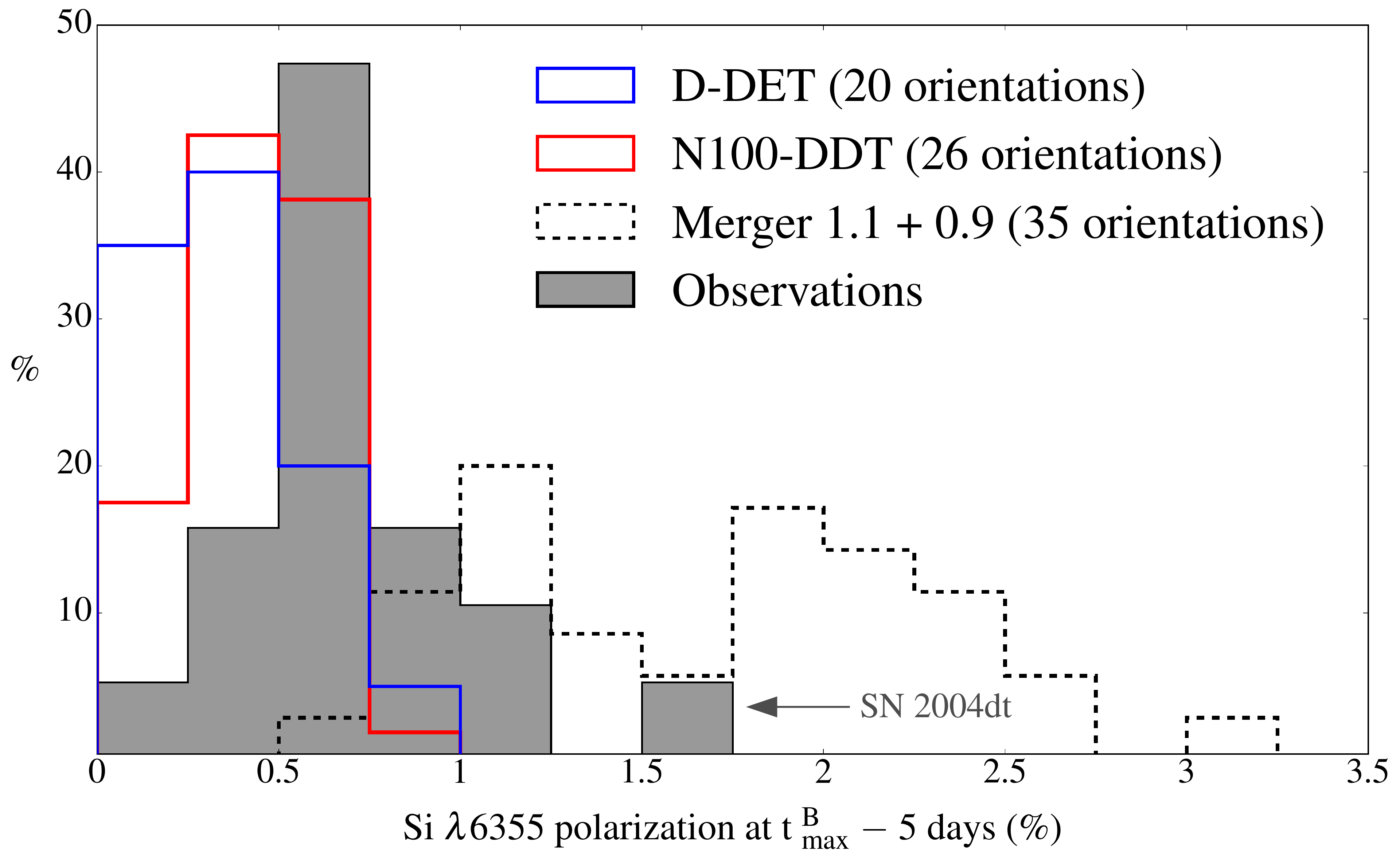}
\caption{Solid-angle-weighted distribution of Si\,{\sc ii}~$\lambda6355$ polarization at 5~d before $B-$band maximum for the \subch (blue distribution) and \ch (red distribution) models, together with the distribution of values observed for normal SNe~Ia \citep[grey distribution,][]{wang2007,patat2009,maund2013,patat2014,porter2016}. The distribution of each explosion model includes orientations from both high and low signal-to-noise calculations (see Section~\ref{radtransf}). For comparison, the dashed purple distribution reports values extracted along 35 different viewing angles for the violent merger model of \citet[][see \citealt{bulla2016}]{pakmor2012}.}
\label{distribution}
\end{center}
\end{figure*}

Spectropolarimetric observations of SN~2004dt were acquired with FORS1 at ESO VLT about a week before maximum \citep{wang2006} and with the Shane telescope at Lick Observatory 4~d after maximum \citep{leonard2005}. Except for the pseudo-continuum range, where degrees of polarization typical of normal SNe~Ia ($\sim$~0.3$-$0.4~per~cent) were detected, SN~2004dt displayed unusual polarization properties. Together with a very strong ($\sim$~2$-$2.5~per~cent) peak around 3650~\AA{}, several features polarized at $\sim$~1~per~cent level were observed in the blue region and attributed to Si\,{\sc ii}, Mg\,{\sc ii} and Fe\,{\sc ii} lines \citep{wang2006}. Furthermore, the Si\,{\sc ii}~$\lambda6355$ line was unusually strong in polarization, with a level of $\sim$~1.7$-$1.8~per~cent at peak (see also Fig.~\ref{distribution}). Taken together, these properties make SN~2004dt the \revised{most polarized} SN~Ia ever observed.

Fig.~\ref{2004dt} shows pre- and post-maximum polarization spectra of SN~2004dt together with those extracted for the two explosion models studied in this paper. The agreement in the flux spectra of the models with SN~2004dt is of comparable quality to that with SN~2001el and SN~2012fr. However, we clearly see that none of the orientations investigated for the \subch and \ch model can reproduce the degrees of polarization observed in SN~2004dt. In contrast, a better match to SN~2004dt was found by \citet{bulla2016} for the least polarized line of sights of the violent merger of \citet{pakmor2012}. Although both explosion models studied in this paper reproduce the flux spectra comparably well to the violent merger model, spectropolarimetry does provide a very clear distinction. Specifically, for both models, the synthetic spectropolarimetry hardly ever exceeds a level of ~0.5 per cent, which is clearly insufficient to account for the exceptional polarization levels of SN~2004dt.  


\subsection{Polarization of the Si\,{\sc ii}~$\lambda6355$ feature}
\label{polsilicon}

Si\,{\sc ii}~$\lambda6355$, one of the hallmark features in the spectra of SNe~Ia, is generally found to be polarized. Around maximum light the degree of polarization is typically below 1~per~cent, although larger values are also found \citep[see][]{wang2007}. In this section, we compare the distribution of values observed in normal SNe~Ia with those predicted in this study for the \subch and \ch model. Observed values include 16 normal SNe~Ia from \citet{wang2007}, SN~2006X from \citet{patat2009}, SN~2012fr from \citet{maund2013} \revised{and SN~2014J from \citet{patat2014} and \citet{porter2016}}. To facilitate the comparison with data, we follow the convention of \citet{wang2007} and present synthetic polarization values estimated via Gaussian fitting of the Si\,{\sc ii}~$\lambda6355$ profile at 5~d before $B-$band maximum light. To explore the full range of polarization covered by each model, here we also include predictions from our low signal-to-noise calculations with spectra extracted for an additional 17 (\mbox{D-DET}) and 21 (\mbox{N100-DDT}) orientations (see Section~\ref{radtransf}). 
 
Fig.~\ref{distribution} shows the solid-angle-weighted distribution of Si\,{\sc ii}~$\lambda6355$ polarization values extracted for our models compared to those detected in normal SN~Ia events. This comparison clearly demonstrates that our models produce polarization signals in a range ($\lesssim$~1~per~cent) comparable to the values observed. Although the range of polarization predicted by the two models is relatively similar, the \subch model typically produces lower levels of polarization than the \ch model (in line with the findings of Section~\ref{synthobs}) and it provides a good match with observed SNe at the low-polarization end of the distribution. Nonetheless, neither the \subch nor the \ch model can account for the
relatively high Si\,{\sc ii}~$\lambda6355$ polarization levels ($\gtrsim$~1~per~cent) observed in a handful of SNe~Ia \citep{wang2007,patat2009}. We note, however, that polarization signals for objects at the upper end of the distribution are associated with high-velocity components of the Si\,{\sc ii}~$\lambda6355$ profile. As none of our explosion models is able to produce high-velocity features in the Si\,{\sc ii}~$\lambda6355$ line, it is not surprising that the match in polarization with these relatively highly-polarized objects is also poor. 

For comparison, in Fig.~\ref{distribution} we also include polarization values extracted for the violent merger model of \citet{pakmor2012} along 35 different orientations \citep{bulla2016}. As discussed by \citet{bulla2016}, compared to observations, the silicon line is typically too highly polarized in this model, with levels above \revised{1.75~per~cent for about half of the orientations}. However, predictions for the remaining orientations match the polarization observed for ``highly''-polarized events ($\gtrsim$~1~per~cent) like SN~2004dt \citep[see also fig.~11 of][]{bulla2016}.

\subsection{Polarization of the O\,{\sc i}~/~Si\,{\sc ii}~/~Mg\,{\sc ii} feature}
\label{poloxygen}

\begin{figure*}
\begin{center}
\includegraphics[width=0.97\textwidth]{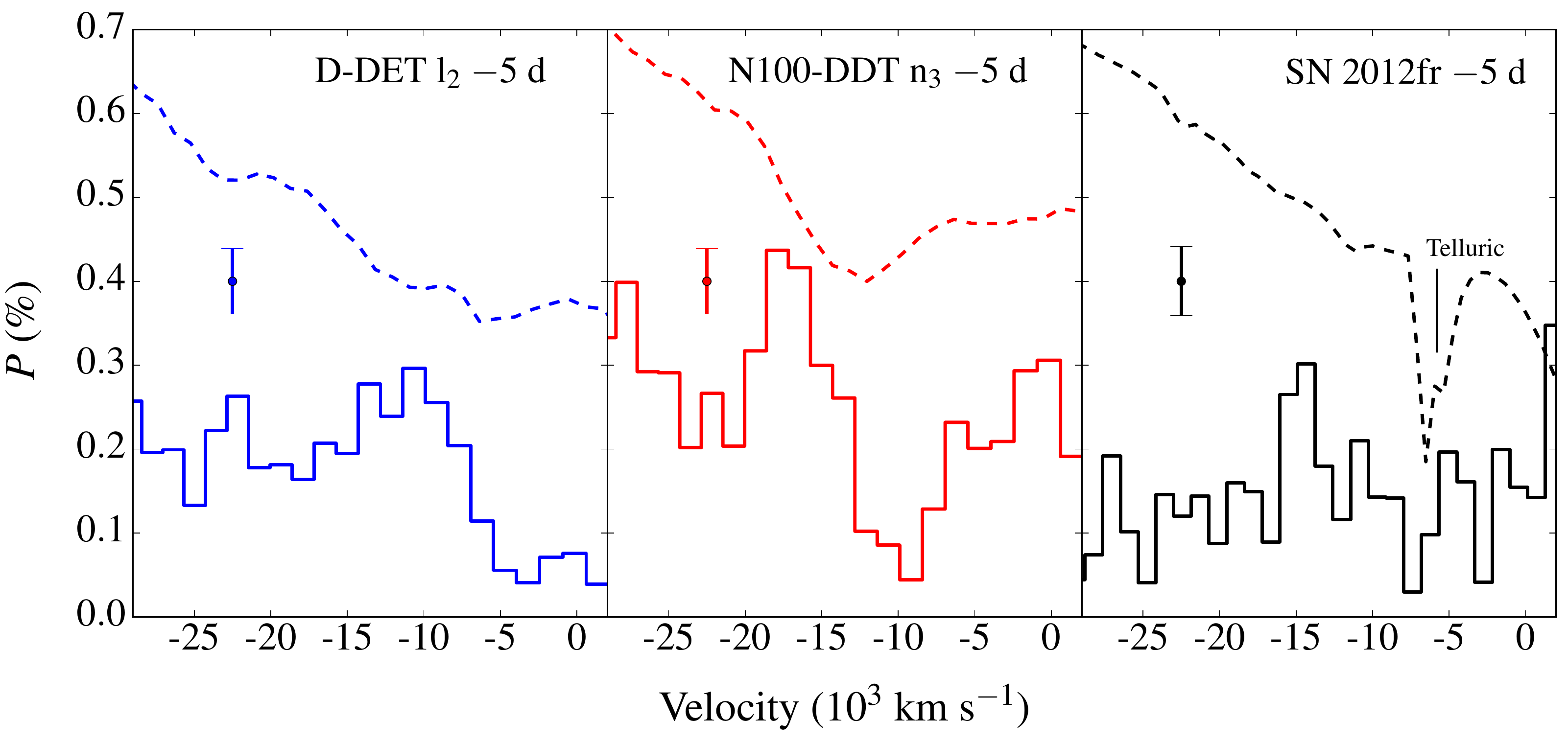}
\caption{Polarization spectra (solid lines) around the O\,{\sc i}~$\lambda7774$ feature at 5~d before maximum. Spectra extracted for the \subch $\bmath{l_2}$ (left-hand panel) and the \ch $\bmath{n_3}$ (middle panel) orientations are reported together with those observed for SN~2012fr (right-hand panel). Flux spectra are arbitrarily scaled for presentation and reported with dashed lines in each panel. Error bars mark the uncertainties on the synthetic (left-hand and middle panels) and observed (right-hand panel) polarization spectra.}
\label{oxygen}
\end{center}
\end{figure*}

Unlike the cases of Si\,{\sc ii} and Ca\,{\sc ii}, no unambiguous polarization signature has ever been observed for the O\,{\sc i} feature in SNe~Ia \citep{wang2008}. In particular, even when investigating relatively polarized objects as SN~2004dt and SN~2006X, \citet{wang2006} and \cite{patat2009} were only able to place upper limits of $\sim$~0.3~per~cent on the O\,{\sc i}~$\lambda7774$ degree of polarization. \revised{Based on the spectropolarimetric data of SN~2004dt,} \citet{wang2006} have interpreted the fact that silicon, calcium and oxygen features are seen at similar velocities, but with different polarimetric properties, as evidence that aspherically-distributed IMEs should spatially coexist with the more spherically-distributed oxygen layer. \revised{This interpretation was supported by predictions from delayed-detonation models deformed into ellipsoidal geometries \citep{howell2001,hoeflich2006}.} As shown in Fig.~\ref{ejecta}, the distribution of oxygen, calcium and silicon in the ejecta of our models is consistent with this interpretation. Indeed, as expected, the more spherical distribution of oxygen, compared to those of silicon and calcium, results in weaker polarization levels across the O\,{\sc i}~$\lambda7774$ region than across the Si\,{\sc ii} and Ca\,{\sc ii} lines.

Despite being weak, however, polarization signatures \revised{in the} O\,{\sc i}~$\lambda7774$ \revised{wavelength region} are clearly discernible in both our models and \revised{may} be detectable with slightly more sensitive spectropolarimetric observations. \revised{In particular, in Section~\ref{geometry} we have seen that the polarization peaks around this region can be associated with either O\,{\sc i}~$\lambda7774$, Si\,{\sc ii}~$\lambda7849$, Mg\,{\sc ii}~$\lambda7887$ or a mixture of these three lines. In Fig.~\ref{oxygen} we show a zoom into the O\,{\sc i}~$\lambda7774$ region for the observed spectra of SN~2012fr and the synthetic spectra predicted by our models. Spectra are reported at 5~d before maximum and are the same as in the left-hand panels of Fig.~\ref{2012fr}. Polarization peaks at the 0.3 and 0.4~per~cent level are predicted across this spectral region by both explosion models. A similar analysis to that presented in Section~\ref{geometry} suggests that the polarization signal across this feature is dominated by O\,{\sc i}~$\lambda7774$ for both these orientation/model/epoch combinations, and the predicted
velocities for this feature are $\sim-$10\,000~km~s$^{-1}$ for the \subch and $\sim-$17\,000~km~s$^{-1}$ for the \ch model.} Although \citet{maund2013} did not discuss any possible signal associated with O\,{\sc i}~$\lambda7774$, we notice a marginal increase in the polarization levels up to $\sim$~0.2$-$0.3~per~cent between $-$10\,000 and $-$15\,000~km~s$^{-1}$ that, owing to the resemblance with the predicted profiles, might be associated with O\,{\sc i}~$\lambda7774$.

\revised{Overall, we conclude that the spectral region around the O\,{\sc i} / Si\,{\sc ii} / Mg \,{\sc ii} feature is an interesting target for future spectropolarimetric observations since the models predict that weak signatures may be present here. In addition, predictions from the \subch model suggest that these signatures may be observed rotated of 90$^\circ$ with respect to Si\,{\sc ii}~$\lambda6355$, Ca\,{\sc ii}~IR and other lines in the spectrum.  However, we caution that interpretation will be complicated by the competing roles of the O\,{\sc i}, Si\,{\sc ii} and Mg \,{\sc ii} spectral features in this range for which the relative contributions is likely to vary as a function of epoch and observer orientation, as highlighted by the discussion in Section~\ref{geometry}.} 

\subsection{Polarization of the Ca\,{\sc ii} IR triplet feature}
\label{polcalcium}

\revised{As shown in Section~\ref{compsn}, our two explosion models fail to reproduce the polarization signatures observed in SNe~Ia across the Ca\,{\sc ii} IR triplet feature. The disagreement is particularly evident across the high-velocity ($\sim$~15\,000$-$20\,000~km~s$^{-1}$) components of this line that are frequently detected in both intensity (e.g. \citealt{gerardy2004}; \citealt{mazzali2005a}a; \citealt{childress2014,maguire2014,silverman2015,zhao2015}) and polarization \citep[e.g.][]{wang2003,maund2013} spectra of SNe~Ia.}

\revised{In the \ch model, typical values for the velocities ($\sim$12\,000$-$15\,000~km~s$^{-1}$) and the polarization signals ($\sim$~0.3~per~cent) extracted across the Ca\,{\sc ii} IR triplet line are too low to match the observations, as highlighted by the spectral comparison with SN~2001el (Fig.~\ref{2001el}) and SN~2012fr (Fig.~\ref{2012fr}). This comes as no surprise, since the calcium distribution in the ejecta of this model extends up to $\sim$~15\,000~km~s$^{-1}$ and is not strongly asymmetric (see Fig.~\ref{ejecta}).}

\revised{The \subch model appears more promising in matching the calcium high-velocity components since the distribution of calcium in this model can reach up to $\sim$~30\,000~km~s$^{-1}$ for orientations close to the north pole. Indeed, we find a reasonable match between velocities predicted for the \subch model along the equatorial direction ($\bmath{l_2}$) and those observed in both SN~2001el (Fig.~\ref{2001el}) and SN~2012fr (Fig.~\ref{2012fr}). This orientation, however, is associated with polarization signals across the calcium line that are inconsistent with data. Because of the strong asymmetries in the outer shell of this model (see Fig.~\ref{ejecta}), selecting orientations out of the equatorial plane typically leads to poorer matches with observations: calcium lines for viewing angles in the southern hemisphere are typically too weakly polarized and too slow, while those in the northern hemisphere too strongly polarized and too fast (see polarization spectra around maximum light in Fig.~\ref{specmax_subch}).}

\revised{Although high-velocity features in the Ca\,{\sc ii} IR triplet are frequently observed in SN~Ia spectra, their origin remains unclear and may be associated with either a density increase (\citealt{gerardy2004}; \citealt{mazzali2005b}b; \citealt{tanaka2006}), a composition enhancement (\citealt{mazzali2005b}b) or a ionization effect \citep{blondin2013} in the outer ejecta. In the single-degenerate scenario, a density increase might be associated with circumstellar material (\citealt{gerardy2004}; \citealt{mazzali2005b}b; \citealt{tanaka2006}). If that is the case, the lack of high-velocity features in the \ch model would not be surprising as circumstellar material is not captured by the \citet{seitenzahl2013} explosion models. Alternatively, an abundance enhancement in the outer regions of the ejecta (\citealt{mazzali2005b}b) would seem qualitatively consistent with the calcium distribution predicted in the He-shell by our \subch model. As noted above, however, the \subch model struggles to simultaneously reproduce spectroscopic and polarimetric properties of this high-velocity material. In the future development of double-detonation models, it will be then particularly important to consider whether better agreement in the high-velocity Ca~{\sc ii} features can be achieved by models in which properties of the ash of the He-shell detonation (both composition and geometry) differ from the predictions of the \citet{fink2010} simulations. In particular, we note that it has already been suggested that changes in the He-shell properties are likely required if such models are to provide a good match to observed optical light curves and colours \citep[see e.g.][]{kromer2010}. It remains to be seen whether multi-dimensional explosion models with altered He shell properties can produce morphologies more consistent with the observed spectropolarimetry of high-velocity features. }

\section{Discussion and Conclusions}
\label{conclusions}

We have presented polarization calculations for one double-detonation model from \citet[][here referred to as D-DET]{fink2010} and one delayed-detonation model from \citet[][here N100-DDT]{seitenzahl2013}. Using the techniques of \citet{bulla2015}, we calculated polarization spectra around maximum light and focused on three orientations for the 2D \subch model and five for the 3D \ch model. In order to map out the range of polarization covered by the models, we also performed extra-calculations that provide lower signal-to-noise spectra for an additional 17 (\mbox{D-DET}) and 21 (\mbox{N100-DDT}) viewing angles.

For both models, the overall polarization levels are low at all the epochs considered in this study (between 10 and 30~d after explosion): they are consistently below 1~per~cent, with the largest values generally observed for the \ch model. While the 2D \subch model is axi-symmetric by construction, polarization spectra for the 3D \ch model verify that polarimetry could be used to identify deviations from a single-axis geometry. A common behaviour between the two explosion models is the relative rise in polarization level from blue to red wavelengths. Our calculations confirm that this effect, which is regularly observed in both normal \citep[e.g.][]{wang1997} and sub-luminous \citep[e.g.][]{howell2001} SNe~Ia, can be ascribed to the decreasing line blanketing when moving to longer wavelengths, and is fairly well captured in the explosion models: in our simulations around maximum light, the fraction of escaping packets having a depolarizing line interaction as last event drops from about 0.8 at 4000~\AA{} to only 0.3 at 7000~\AA.

Polarization spectra predicted by our models match particularly well those observed for the majority of normal SNe~Ia, as clearly highlighted by the comparison with the two well-studied supernovae SN~2001el and SN~2012fr. Polarization levels in the pseudo-continuum range (between 6500 and 7500~\AA) are found to decrease from $\sim$~0.1$-$0.3~per~cent at maximum light to lower values at later times, in strikingly good agreement with observations. Higher degrees of polarization are predicted at wavelengths corresponding to the troughs of absorption lines. In particular, the set of polarization signals extracted across the Si\,{\sc ii}~$\lambda6355$ feature ($\lesssim$~1~per~cent) provides a remarkable match with the distribution of values observed in normal SNe~Ia \citep{wang2007,patat2009,maund2013}, while the low degrees of polarization found in the O\,{\sc i}~$\lambda7774$ region are consistent with the non-detection of this feature in currently available data. Taken together, these findings lead us to conclude that the geometries of both the explosion models considered here are likely to be broadly consistent with those of normal SNe~Ia.

However, none of our models is able to reproduce the high-velocity ($\sim$~15\,000$-$20\,000~km~s$^{-1}$) components of the Ca\,{\sc ii} IR triplet and Si\,{\sc ii}~$\lambda6355$ lines that are frequently observed in both intensity (e.g. \citealt{gerardy2004}; \citealt{mazzali2005a}a; \citealt{childress2014,maguire2014,silverman2015,zhao2015}) and polarization \citep[e.g.][]{wang2003,leonard2005,wang2006,maund2013} spectra of SNe~Ia. Although the \subch scenario is more promising in this sense -- as the calcium in the outer shell of this model leads to velocities in the right range for some orientations -- none of our models is able to simultaneously produce the spectroscopic and polarimetric signatures characteristic of this high-velocity material. The failure to predict silicon and calcium in the right location and with the right geometry clearly indicates that more work is needed to understand whether the two scenarios investigated in this study can account for the high-velocity features observed in the Si\,{\sc ii}~$\lambda6355$ and Ca\,{\sc ii} IR triplet lines.

The two models presented in this paper are hard to distinguish based on their polarization signatures. 
In both, polarizing electron scattering contributions originate in rather spherical regions of the ejecta (between $\sim$~5000 and 15\,000~km~s$^{-1}$), thus leading to polarization levels that are typically low and similar for the two models at most wavelengths.
It is important to note, however, that we have only calculated polarization signatures for two specific models and the results presented here are not necessarily representative of the whole \citet{fink2010} double-detonation and \citet{seitenzahl2013} DDT studies. While the different double-detonation models of \citet{fink2010} are all characterized by relatively similar geometries -- and are thus expected to produce comparable polarization signals -- the DDT models of \citet{seitenzahl2013} yield ejecta whose morphologies are strongly dependent on the ignition configuration (see also Section~\ref{chmodel}). In particular, compared to the N100 model used in this paper, models in which the explosion is initiated with a smaller number of kernels (1 to 40) are typically more asymmetric and thus likely to be more polarized. In the future, it will be crucial to map out the polarization range spanned by the full set of models of \citet{fink2010} and \citet{seitenzahl2013} to understand whether the two scenarios can be effectively distinguished via variations between the polarization signatures across the model sequences.

This study is part of a long-term project aimed to test multi-dimensional hydrodynamic explosion models via their polarization signatures. In a previous paper \citep{bulla2016}, we have demonstrated that the violent merger of a 1.1 and a 0.9~M$_{\odot}$ WD \citep{pakmor2012} is unlikely to explain polarization features observed for the bulk of normal SNe~Ia. In contrast, this work shows that the realizations of the double-detonation and DDT scenarios studied here produce polarization levels in much better agreement with spectropolarimetric data of normal SNe~Ia. To investigate whether the polarization properties of the discussed model classes are representative for the observed SN~Ia population, in the future we aim to map out the parameter space of all the three explosion scenarios.



\section*{Acknowledgements}

\revised{We thank the anonymous reviewer for his/her valuable comments.} We are grateful to Douglas Leonard and Justyn Maund for providing polarization data of SN~2004dt and SN~2012fr reported in this work. MB thanks Ferdinando Patat for many useful discussions about modelling and observing polarized light from supernovae. 

This work used the DiRAC Complexity system, operated by the University of Leicester IT Services, which forms part of the STFC DiRAC HPC Facility (www.dirac.ac.uk). This equipment is funded by BIS National E-Infrastructure capital grant ST/K000373/1 and STFC DiRAC Operations grant ST/K0003259/1. DiRAC is part of the National E-Infrastructure.

This research was supported by the Partner Time Allocation (Australian National University), the National Computational Merit Allocation and the Flagship Allocation Schemes of the NCI National Facility at the Australian National University. Parts of this research were conducted by the Australian Research Council Centre of Excellence for All-sky Astrophysics (CAASTRO), through project number CE110001020.

The authors gratefully acknowledge the Gauss Centre for Supercomputing
(GCS) for providing computing time through the John von Neumann
Institute for Computing (NIC) on the GCS share of the supercomputer
JUQUEEN \citep{stephan2015} at J\"ulich Supercomputing Centre
(JSC). GCS is the alliance of the three national supercomputing
centres HLRS (Universit\"at Stuttgart), JSC (Forschungszentrum
J\"ulich), and LRZ (Bayerische Akademie der Wissenschaften), funded by
the German Federal Ministry of Education and Research (BMBF) and the
German State Ministries for Research of Baden-W\"urttemberg (MWK),
Bayern (StMWFK) and Nordrhein-Westfalen (MIWF).

SAS acknowledges support from STFC grant ST/L0000709/1. IRS was supported by the Australian Research Council Laureate Grant FL0992131. The work of FKR and RP is supported by the Klaus Tschira Foundation. RP acknowledges support by the European Research Council under ERC-StG grant EXAGAL-308037, WH and ST by the DFG through the Cluster of Excellence
'Origin and Structure of the Universe' and the Transregio Project
33 'The Dark Universe'.

\bibliographystyle{mn2e}
\bibliography{bulla2016b}

\end{document}